\documentclass[a4paper,fleqn]{cas-sc}
\usepackage[numbers]{natbib}
\usepackage{hyperref}
\usepackage[dvipsnames]{xcolor}

\def\tsc#1{\csdef{#1}{\textsc{\lowercase{#1}}\xspace}}
\tsc{WGM}
\tsc{QE}
\tsc{EP}
\tsc{PMS}
\tsc{BEC}
\tsc{DE}

\begin{document}
\let\WriteBookmarks\relax
\def\floatpagepagefraction{1}
\def\textpagefraction{.001}

\shorttitle{Interaction size and dissent tolerance in majority-rule dynamics with collective reversal}

\shortauthors{R.~Muslim, R.~Anugraha~NQZ,  Q.~G.~Haq, F.~Nugroho, I.~S.~Alam, and E.~L.~Istiqomah}

\title[mode = title]{Interaction size and dissent tolerance in majority-rule dynamics with collective reversal}    
\author[1,2]{Roni Muslim}[orcid= 0000-0001-6925-5923]
\ead{roni.muslim@apctp.org}
\author[3]{Rinto Anugraha NQZ}[orcid=0000-0002-4781-4090]
\ead{rinto@ugm.ac.id}
\cormark[1]
\cortext[cor1]{Corresponding author}

\author[3]{Qonuni Gusthaf Haq}
\ead{qonuni.haq@gmail.com}

\author[3]{Fahrudin Nugroho}
\ead{fakhrud@ugm.ac.id}

\author[3]{Idham Syah Alam}
\ead{idham@ugm.ac.id}

\author[3]{Elida Lailiya Istiqomah}
\ead{elida@ugm.ac.id}


\affiliation[1]{
    organization={Asia Pacific Center for Theoretical Physics (APCTP)},
    city={Pohang},
    postcode={37673},
    country={Republic of Korea}
}
    
\affiliation[2]{organization={Research Center for Quantum Physics,
    National Research and Innovation Agency (BRIN)},
    city={South Tangerang},
    postcode={15314}, 
    country={Indonesia}}
    
\affiliation[3]{organization={Department of Physics,
    Gadjah Mada University},
    city={Yogyakarta},
    postcode={55281}, 
    country={Indonesia}}

\begin{abstract}
We introduce a majority-rule model in which collective reversal can be activated in highly aligned groups even when a limited number of members dissent. The dissent tolerance $d$ extends the strict-unanimity dynamics by making near-unanimous group compositions eligible for reversal. Mean-field analysis and simulations reveal that this change qualitatively alters the phase structure. Under strict unanimity, a physically accessible transition exists only for $n=3$ and $n=4$. Allowing dissent restores transitions at larger interaction sizes, replacing the fixed interaction-size threshold with an accessibility boundary in the $(n,d)$ plane. When the activation window is sufficiently broad, directional asymmetry can eliminate one of the two ordered attractors through a saddle-node bifurcation, producing a single stable collective state. In the one-sided case, increasing the dissent tolerance can shorten the
transient approach to consensus but leaves its leading logarithmic dependence
on population size unchanged. Activation selectivity acts as an independent control parameter for collective ordering, bistability, and consensus dynamics.
\end{abstract}





\maketitle

\section{Introduction}

Opinion-dynamics models relate individual interaction rules to collective outcomes such as consensus, coexistence, polarization, and abrupt changes in public opinion
\cite{castellano2009statistical,galam2008sociophysics,
tsintsaris2024dynamics,starnini2025opinion}. Well-known binary-state examples include the voter
\cite{clifford1973model,holley1975ergodic,castellano2009nonlinear},
Sznajd \cite{sznajd2000opinion}, and majority-rule models
\cite{galam2002minority}. Early formulations often assumed pairwise interactions, although many social decisions are made in groups. Models with polyadic or higher-order interactions have since shown that group structure and interaction size can change phase transitions, relaxation, consensus formation, and polarization
\cite{steinmeyer2025polyadic,sampson2025group,
perezmartinez2025polarization,muslim2026random}. Higher-order interactions can also modify the stability and bifurcation structure of collective decision-making processes
\cite{battiston2025higher,njougouo2026collective}. In majority-rule dynamics, a selected group adopts its strict local majority
\cite{galam2008sociophysics,sirbu2017opinion}. This rule represents conformity to the opinion that is locally dominant and tends to amplify small population imbalances. Repeated updates can then drive the population toward either of two symmetry-related ordered states. The outcome depends on the size and composition of the selected groups
\cite{muslim2024impact,mulya2024phase}. When the group size varies, its distribution also affects the stability of collective ordering and the relaxation time
\cite{muslim2026random}. The interaction size is therefore part of the microscopic rule that determines the macroscopic behavior.

Majority reinforcement can be weakened or opposed by nonconformist responses. Contrarian agents, for example, choose the opinion opposite to the local majority and can destabilize an ordered state when their influence is sufficiently strong
\cite{galam2004contrarian}. External fields, independence, noise, zealotry, and directed persuasion have also been used to represent the effects of media, political campaigns, institutions, and other sources of external information
\cite{sirbu2017opinion,colaiori2015interplay,mobilia2003does,
civitarese2021external}. Depending on the update rule, these mechanisms can shift transition points, suppress consensus, stabilize coexistence, or favor one opinion
\cite{mukhopadhyay2020voter,azhari2023external}. Their effect may also vary with the social structure through which the external influence propagates
\cite{vansanten2024clustering}.

The microscopic implementation of these responses is important. Directed propaganda acts by making agents follow a prescribed opinion with a given probability and produce an order--disorder transition for triplet interactions
\cite{forgerini2024directed}. Asymmetric contrarian populations break the exchange symmetry between the two opinions and shift the stationary states toward the favored direction
\cite{galam2019asymmetric}. Biased contrarian responses in nonlinear voter dynamics can similarly generate asymmetric stationary states
\cite{pradhan2026contrarian}. Another formulation assigns different contrarian probabilities to unanimous and mixed-majority triplets
\cite{galam2026ratio}. There, the response is applied at the individual level, and the number of agents switching opinions can vary within a selected group. Collective reversal considered here instead changes the outcome of the group as a correlated unit.

A previous study applied collective reversal to unanimous triplets on quenched networks
\cite{muslim2026topology}. Mixed triplets followed majority rule, while unanimous triplets could reverse to the opposite opinion. Degree heterogeneity and clustering shifted the transition away from the well-mixed prediction, especially on strongly clustered Watts--Strogatz networks. The analysis was restricted to $n=3$, and reversal required perfect unanimity. This requirement becomes restrictive as $n$ grows. In a balanced, well-mixed population, the total binomial weight of the two unanimous configurations is $2^{1-n}$, so their occurrence decreases exponentially with the interaction size. Even a large reversal probability may then have little effect because eligible groups are rarely selected.

The unresolved issue is whether this loss of effectiveness is caused by the interaction size itself or by the strict activation condition. We address it by allowing collective reversal in groups that contain a limited number of dissenters. This separates two aspects of the response that coincide in the unanimity model: the probability that reversal occurs after activation and the range of group compositions in which activation is possible. The latter is referred to as activation selectivity. Higher-order environments with limited exposure to disagreement provide a qualitative setting for such a rule
\cite{perezmartinez2025polarization}. Echo chambers offer a related example in which local information is concentrated around similar views
\cite{quattrociocchi2016echo,cinelli2021echo}, while experimental evidence indicates that responses to agreement and dissent may depend on the prevailing level of consensus
\cite{mellody2025dissent}. These connections motivate the model at a qualitative level; the dissent tolerance is not introduced as a directly measurable psychological quantity.

We consider a well-mixed population updated in groups of fixed size $n$. If $\ell$ is the number of agents holding opinion $+1$, a group is eligible for reversal when $\ell\leq d$ or $\ell\geq n-d$, where
$0\leq d\leq\lfloor(n-1)/2\rfloor$. The integer $d$ specifies the number of dissenters that can be present without preventing activation. Strict unanimity corresponds to $d=0$, whereas $d>0$ includes near-unanimous groups. Eligible groups dominated by opinion $-1$ reverse to $+1$ with probability $\epsilon_{\uparrow}$, and those dominated by $+1$ reverse to $-1$ with probability $\epsilon_{\downarrow}$. Otherwise, the group follows majority rule. The probabilities $\epsilon_{\uparrow}$ and $\epsilon_{\downarrow}$ set the response after an eligible group has been selected, while $d$ determines how often such groups can occur. This activation rule differs from homogeneous external fields, configuration-independent noise, and directed propaganda, which do not require a highly aligned local configuration
\cite{civitarese2021external,azhari2023external,
forgerini2024directed,vansanten2024clustering}. It also differs from ratio-dependent contrarian activation
\cite{galam2026ratio} because the response is correlated across the selected group and controlled by the tolerance $d$. The present formulation complements the network model in Ref.~\cite{muslim2026topology}. Rather than changing the quenched topology at fixed $n=3$ and $d=0$, we examine how $n$ and $d$ interact in a well-mixed population.

Mean-field analysis and Monte Carlo simulations give a stability threshold that depends on both $n$ and $d$. For $d=0$, a symmetric transition is physically accessible only for $n=3$ and $n=4$. Allowing dissent increases the statistical weight of reversal-eligible groups and restores accessible transitions as $n$ increases. The interaction-size cutoff of the unanimity model is replaced by a boundary in the $(n,d)$ plane that specifies the minimum tolerance required for a physical transition. At fixed $d$, reversal again becomes ineffective when $n$ is sufficiently large, whereas a tolerance that broadens with $n$ keeps the transition accessible. Symmetric reversal yields supercritical pitchfork bifurcations in the cases studied. Directional asymmetry breaks this structure through saddle-node bifurcations and can remove one of the competing ordered attractors. In the one-sided limit, changing $d$ affects the transient approach to consensus but leaves the leading logarithmic dependence on population size unchanged. Activation selectivity, therefore, provides a separate control over the stationary and transient dynamics.

\section{Model and methods}
\label{sec:model}

\subsection{Microscopic dynamics}
\label{sec:microscopic_model}

We consider a majority-rule opinion dynamics model with
homogeneity-triggered collective reversal in a well-mixed population of size
$N$. Each individual $i=1,2,\ldots,N$ holds a binary opinion $s_i=\pm1$,
representing two competing alternatives. The microscopic state of the
population is specified by $\{s_1,s_2,\ldots,s_N\}$. At the macroscopic level,
the state is characterized by the fraction $c=N_{+}/N$, where $N_{+}$ is the
number of individuals holding opinion $+1$. The normalized opinion imbalance
is $m=2c-1=N^{-1}\sum_{i=1}^{N}s_i$. Thus, $m=1$ and $m=-1$ correspond to
the two consensus states, whereas $m=0$ represents equal macroscopic fractions
of the two opinions.

The model extends Galam's majority-rule dynamics
\cite{galam2002minority,galam2008sociophysics} by introducing a collective
response that may reverse the outcome of a highly homogeneous group. The
activation condition is controlled by an integer dissent tolerance $d$, defined
as the maximum number of agents holding the local minority opinion for which
collective reversal remains possible. Its admissible range is
$0\leq d\leq\lfloor(n-1)/2\rfloor$, where $n$ is the interaction-group size.
Equivalently, the normalized homogeneity threshold is
$\rho_d=1-d/n$. A selected group is eligible for collective reversal when at
least a fraction $\rho_d$ of its members hold the same opinion. The case $d=0$ corresponds to perfect unanimity: collective reversal is
activated only when all selected individuals initially agree. For $d>0$, a
limited number of dissenters is tolerated, so that near-unanimous groups may
also undergo collective reversal. At the opposite limit,
$d=\lfloor(n-1)/2\rfloor$, every group with a strict local majority becomes
eligible. The parameter $d$ therefore interpolates between reversal restricted
to perfect unanimity and collective anticonformity acting on the complete
strict-majority sector.

At each elementary update, $n$ distinct individuals are selected uniformly at
random from the population. Because the population is well mixed, there is no
underlying spatial or network structure, and all individuals are statistically
equivalent. Let $\ell$ denote the number of selected individuals holding
opinion $+1$; the remaining $n-\ell$ individuals hold opinion $-1$. The
microscopic dynamics is defined as follows:

\begin{enumerate}
    \item If $\ell=n/2$, which is possible only for even $n$, the selected
    group is equally divided, and no update occurs.

    \item If $\ell<n/2$, opinion $-1$ constitutes the strict local majority.

    \begin{enumerate}
        \item If $\ell\leq d$, the group satisfies the homogeneity condition.
        With probability $\epsilon_{\uparrow}$, the group undergoes collective
        reversal and all its members adopt opinion $+1$. With the complementary
        probability $1-\epsilon_{\uparrow}$, majority rule is applied and all
        group members adopt opinion $-1$.

        \item If $d<\ell<n/2$, the group does not satisfy the homogeneity
        condition and evolves only through majority rule, so all group members
        adopt opinion $-1$.
    \end{enumerate}

    \item If $\ell>n/2$, opinion $+1$ constitutes the strict local majority.

    \begin{enumerate}
        \item If $\ell\geq n-d$, the group satisfies the homogeneity condition.
        With probability $\epsilon_{\downarrow}$, the group undergoes
        collective reversal and all its members adopt opinion $-1$. With
        probability $1-\epsilon_{\downarrow}$, majority rule is applied and
        all group members adopt opinion $+1$.

        \item If $n/2<\ell<n-d$, the group does not satisfy the homogeneity
        condition and evolves only through majority rule, so all group members
        adopt opinion $+1$.
    \end{enumerate}
\end{enumerate}
Here, collective reversal means that the final group opinion is opposite to its
initial local majority. For a unanimous group, this requires every selected
individual to change their opinion. For a near-unanimous group, agents initially in
the minority already hold the final reversal opinion and therefore remain
unchanged, while the initial local majority switches collectively. Thus, the
term refers to the reversal of the group-level outcome rather than necessarily to
the simultaneous flipping of every individual.

For $d=0$, majority rule and collective reversal act on distinct local
configurations. Non-unanimous strict-majority groups follow majority rule,
whereas the two unanimous configurations may reverse collectively. For
$d>0$, the two mechanisms compete directly within highly homogeneous groups: majority rule reinforces the local majority, while collective reversal drives the group toward the opposite opinion. Pure majority-rule dynamics is recovered for $\epsilon_{\uparrow}=\epsilon_{\downarrow}=0$, irrespective of $d$. The probabilities $\epsilon_{\uparrow}$ and $\epsilon_{\downarrow}$ determine
the strength and directional asymmetry of collective reversal. The symmetric
case $\epsilon_{\uparrow}=\epsilon_{\downarrow}$ is invariant under the
exchange $+1\leftrightarrow-1$. If
$\epsilon_{\uparrow}>\epsilon_{\downarrow}$, the collective response favors
opinion $+1$, whereas $\epsilon_{\uparrow}<\epsilon_{\downarrow}$ favors
opinion $-1$. In the one-sided limit $\epsilon_{\downarrow}=0$, the all-$+1$
configuration remains absorbing; similarly, the all-$-1$ configuration remains
absorbing when $\epsilon_{\uparrow}=0$.

\begin{figure}[pos = tb!]
    \centering
    \includegraphics[width=0.9\linewidth]{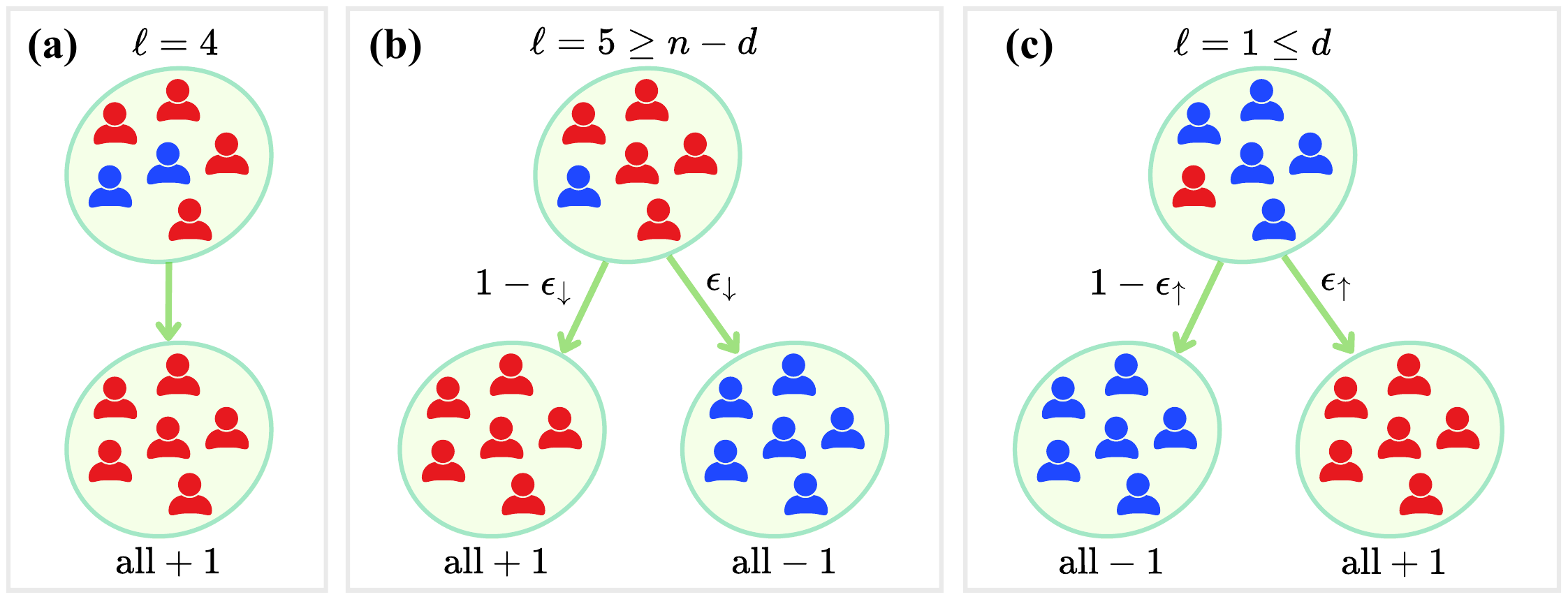}
    \caption{Schematic update rules for the illustrative case $n=6$ and
    $d=1$, with red and blue agents representing opinions $+1$ and $-1$,
    respectively. (a) A non-eligible group with a strict positive majority
    ($\ell=4$) follows majority rule and becomes all-$+1$. (b) An eligible
    positive-majority group ($\ell=5\geq n-d$) becomes all-$+1$ with probability
    $1-\epsilon_{\downarrow}$ or reverses to all-$-1$ with probability
    $\epsilon_{\downarrow}$. (c) An eligible negative-majority group
    ($\ell=1\leq d$) becomes all-$-1$ with probability
    $1-\epsilon_{\uparrow}$ or reverses to all-$+1$ with probability
    $\epsilon_{\uparrow}$. An equally divided group, $\ell=n/2$, undergoes no
    update and is not shown.}
    \label{fig:diagram}
\end{figure}

The parameter $d$ controls the number of local configurations on which
collective reversal can act. At the mixed population state, the probability of
selecting a reversal-eligible group increases from $2^{1-n}$ at $d=0$ to
$2^{1-n}\sum_{\ell=0}^{d}\binom{n}{\ell}$ for general $d$. Consequently, allowing dissent creates more opportunities for collective
reversal, although the actual event frequency also depends on
$\epsilon_{\uparrow}$ and $\epsilon_{\downarrow}$.

The two update channels have opposite effects. Majority rule strengthens the opinion already dominant in the selected group, whereas collective reversal redirects an eligible group toward the opposite opinion. The first channel represents conformity arising from normative or informational influence
\cite{asch1955opinions,deutsch1955study,cialdini2004social}. The second may be viewed as a coordinated response to a strongly aligned group, such as a counternarrative or an institutional intervention. This interpretation is only qualitative. The model does not reproduce the details of persuasion in real social settings; its purpose is to examine how an activation threshold and a correlated group response shape the population dynamics.

\subsection{Mean-field formulation}
\label{sec:mean_field_model}

In a well-mixed population (corresponding to a fully-connected network), the composition of the selected group is determined
by the instantaneous global fraction $c$. Under independent sampling, the
probability of selecting a group containing exactly $\ell$ individuals in
state $+1$ is
$P_n(\ell\mid c)=\binom{n}{\ell}c^\ell(1-c)^{n-\ell}$. For sampling without
replacement from a finite population, the exact group-composition probability
is hypergeometric. The binomial form is recovered in the limit $N\gg n$ and is
used throughout the MF analysis. Let $\mathcal{R}_{n,d}(c)$ and $\mathcal{L}_{n,d}(c)$ denote the expected
numbers of individuals entering and leaving state $+1$, respectively, during
one elementary update. The gain and loss rates of the model are 
\begin{align}
    \mathcal{R}_{n,d}(c)
    ={}&
    \epsilon_{\uparrow}
    \sum_{\ell=0}^{d}
    (n-\ell)P_n(\ell\mid c)+
    \sum_{\ell=\lfloor n/2\rfloor+1}^{n-d-1}
    (n-\ell)P_n(\ell\mid c)+
    (1-\epsilon_{\downarrow})
    \sum_{\ell=n-d}^{n-1}
    (n-\ell)P_n(\ell\mid c),
    \label{eq:gamma_u} \\
    \mathcal{L}_{n,d}(c)
    ={}&
    (1-\epsilon_{\uparrow})
    \sum_{\ell=1}^{d}
    \ell P_n(\ell\mid c)+
    \sum_{\ell=d+1}^{\lfloor(n-1)/2\rfloor}
    \ell P_n(\ell\mid c)+
    \epsilon_{\downarrow}
    \sum_{\ell=n-d}^{n}
    \ell P_n(\ell\mid c).
    \label{eq:gamma_d}
\end{align}
An empty sum is understood to be zero. This convention covers all admissible
values of $d$, including the limiting case in which every strict-majority configuration is eligible for collective reversal. The first term in Eq.~\eqref{eq:gamma_u} describes reversal of a group
initially dominated by opinion $-1$. The second term represents ordinary majority-rule updates in groups dominated by opinion $+1$ but not sufficiently homogeneous to activate reversal. The third term describes the majority rule outcomes in positively dominated groups that satisfy the homogeneity condition but do not reverse. The three terms in Eq.~\eqref{eq:gamma_d} have the
corresponding interpretations under the exchange
$+1\leftrightarrow-1$.

A more transparent representation is obtained by separating the pure
majority-rule drift from the collective-reversal correction. The pure majority-rule contribution is
\begin{align}
    M_n(c)
    ={}&
    \sum_{\ell=\lfloor n/2\rfloor+1}^{n-1}
    (n-\ell)P_n(\ell\mid c)-
    \sum_{\ell=1}^{\lfloor(n-1)/2\rfloor}
    \ell P_n(\ell\mid c).
    \label{eq:MR_term}
\end{align}
We define the probabilities of selecting highly homogeneous groups dominated
by opinions $-1$ and $+1$ as
$A_{n,d}^{-}(c)=\sum_{\ell=0}^{d}P_n(\ell\mid c)$ and
$A_{n,d}^{+}(c)=\sum_{\ell=n-d}^{n}P_n(\ell\mid c)$, respectively. These
quantities satisfy $A_{n,d}^{+}(c)=A_{n,d}^{-}(1-c)$.

Let $t$ denote the number of elementary updates, and introduce the rescaled time
$\tau=t/N$. Because a single update changes the global concentration by
$\Delta c=\Delta N_{+}/N$, the deterministic MF evolution is governed
by
\begin{equation}
    \frac{dc}{d\tau}
    =
    v_{n,d}(c)
    =
    M_n(c)
    +
    n\epsilon_{\uparrow}A_{n,d}^{-}(c)
    -
    n\epsilon_{\downarrow}A_{n,d}^{+}(c),
    \label{eq:model_drift}
\end{equation}
which is equivalent to
$v_{n,d}(c)=\mathcal{R}_{n,d}(c)-\mathcal{L}_{n,d}(c)$. Thus,
Eq.~\eqref{eq:model_drift} describes the deterministic temporal evolution of
the macroscopic concentration $c$. Its compact form shows that collective
reversal modifies the pure majority-rule drift through the probabilities of
sampling groups within the two eligible composition sectors. The factors of $n$ in Eq.~\eqref{eq:model_drift} arise from the difference
between the outcomes of majority rule and collective reversal. Consider an
eligible negatively dominated group with $\ell\leq d$. Under majority rule,
the group becomes all-$-1$, producing a change $\Delta N_{+}=-\ell$. If
collective reversal is activated, the same group instead becomes all-$+1$,
giving $\Delta N_{+}=n-\ell$. The difference between these two outcomes is
therefore $(n-\ell)-(-\ell)=n$, independently of the initial group
composition $\ell$. Consequently, the correction to the pure majority-rule
drift for this configuration is
$n\epsilon_{\uparrow}P_n(\ell\mid c)$. Summing over all eligible negatively
dominated configurations produces
$n\epsilon_{\uparrow}A_{n,d}^{-}(c)$. By the same argument, reversal of
eligible positively dominated groups produces the correction
$-n\epsilon_{\downarrow}A_{n,d}^{+}(c)$.

As a special case, perfect unanimity is recovered by setting $d=0$, for which
$A_{n,0}^{-}(c)=(1-c)^n$ and $A_{n,0}^{+}(c)=c^n$.
Equation~\eqref{eq:model_drift} then reduces to
$
v_{n,0}(c)
=
M_n(c)
+n\epsilon_{\uparrow}(1-c)^n
-n\epsilon_{\downarrow}c^n$,
corresponding to the collective-reversal mechanism previously studied in
Ref.~\cite{muslim2026topology}. Under symmetric reversal,
$\epsilon_{\uparrow}=\epsilon_{\downarrow}=\epsilon$, the dynamics remains invariant under the exchange $c\leftrightarrow1-c$. Increasing $d$ broadens the activation region from the two strictly unanimous configurations to finite binomial tails near the ordered boundaries. Physically, this relaxation makes collective reversal less localized in group-composition space, enabling it to compete more effectively with the ordering tendency induced by majority rule. The resulting MF equation becomes asymptotically accurate for fixed $n$ and $d$ in the limit of a sufficiently large well-mixed population, $N\gg n$.

\subsection{Monte Carlo simulation protocol}
\label{sec:mc_method}

Monte Carlo simulations were performed by implementing the microscopic update
rules directly in finite well-mixed populations. At each elementary update,
$n$ distinct individuals were sampled uniformly without replacement, and the
number $\ell$ of individuals holding opinion $+1$ was determined. For an
eligible negatively dominated group, $\ell\leq d$, all members adopted $+1$
with probability $\epsilon_{\uparrow}$ and $-1$ otherwise. For an eligible
positively dominated group, $\ell\geq n-d$, all members adopted $-1$ with
probability $\epsilon_{\downarrow}$ and $+1$ otherwise. Noneligible
strict-majority groups followed their local majority, whereas tied groups
remained unchanged. One Monte Carlo step (MCS) consisted of $N$ elementary updates, so the time measured in MCS coincides with the rescaled time $\tau=t/N$ in Eq.~\eqref{eq:model_drift}.

Stationary-state simulations were performed under symmetric reversal,
$\epsilon_{\uparrow}=\epsilon_{\downarrow}=\epsilon$, for
$(n,d)=(5,0),(5,1),(8,1),$ and $(8,2)$ at population size $N=10^5$. The
probability interval $0\leq\epsilon\leq1$ was sampled using 101 uniformly
spaced values, supplemented by the exact analytical thresholds
$\epsilon_c(5,1)=7/20$ and $\epsilon_c(8,2)=19/42$ and additional points in their neighborhoods. Each realization was initialized in the all-$+1$ state, equilibrated for $2\times10^3$ MCS, and subsequently measured for $10^4$ MCS, with the magnetization recorded once per MCS. Because the symmetric dynamics is invariant under $+1\leftrightarrow-1$, and the measured observable is $|m|$, initialization in the opposite ordered state produces symmetry-related statistics.

The stationary order parameter was defined as $M=\langle|m|\rangle$, where
the average was taken first over the stationary measurement interval and then
over $R=32$ independent realizations. For realization $r$, the time-averaged
value was $M_r=T_{\mathrm{m}}^{-1}\sum_{q=1}^{T_{\mathrm{m}}}|m_r(q)|$, with
$T_{\mathrm{m}}=10^4$, and the reported result was
$M=R^{-1}\sum_{r=1}^{R}M_r$. Statistical uncertainties were calculated as
standard errors of the independent realization averages, thereby avoiding the
treatment of successive measurements within one trajectory as statistically
independent. Independent random-number streams were generated from a fixed
base seed, and error bars were omitted when smaller than the corresponding
symbol size.

Consensus-time simulations were performed under the one-sided condition
$\epsilon_{\downarrow}=0$, for which the all-$+1$ configuration is absorbing.
All trajectories were initialized at $c_0=0.8$, and the stochastic consensus
time was defined as $\tau_{\mathrm{abs}}=t_{\mathrm{abs}}/N$, where
$t_{\mathrm{abs}}$ is the number of elementary updates required to reach the
all-$+1$ state. The representative parameter pairs
$(n,d)=(5,0),(5,1),(8,1)$, and $(8,2)$ were considered, with the upward
reversal probability fixed at $\epsilon_{\uparrow}=0.5$. The population size was varied over $N=10^2,10^3,10^4,10^5,$ and $10^6$. For each parameter set, $10^3$ independent first-passage trajectories were generated, with a maximum
observation time of $5\times10^4$ MCS. The reported consensus times are
ensemble means, and their statistical uncertainties are standard errors
calculated over the independent trajectories. All trajectories reached the
absorbing state within the prescribed observation window.

\section{Results and discussion}
\label{sec:result}

\subsection{Macroscopic drift and symmetry structure}
\label{sec:aggregated_description}

The deterministic evolution of the positive-opinion fraction is governed by
$dc/d\tau=v_{n,d}(c)=\mathcal{R}_{n,d}(c)-\mathcal{L}_{n,d}(c)$, where the
gain and loss rates are given by Eqs.~\eqref{eq:gamma_u} and
\eqref{eq:gamma_d}, respectively. To separate the average reversal strength
from its directional asymmetry, we introduce
$\bar{\epsilon}=(\epsilon_{\uparrow}+\epsilon_{\downarrow})/2$ and
$\Delta\epsilon=(\epsilon_{\uparrow}-\epsilon_{\downarrow})/2$, such that
$\epsilon_{\uparrow}=\bar{\epsilon}+\Delta\epsilon$ and
$\epsilon_{\downarrow}=\bar{\epsilon}-\Delta\epsilon$. Here,
$\bar{\epsilon}$ quantifies the direction-independent reversal strength,
whereas $\Delta\epsilon$ determines its directional bias. In particular,
$\Delta\epsilon>0$ favors opinion $+1$, $\Delta\epsilon<0$ favors opinion
$-1$, and $\Delta\epsilon=0$ corresponds to symmetric reversal, for which neither opinion is preferentially selected. The probability constraints
$0\leq\epsilon_{\uparrow},\epsilon_{\downarrow}\leq1$ imply
$0\leq\bar{\epsilon}\leq1$ and
$|\Delta\epsilon|\leq\min(\bar{\epsilon},1-\bar{\epsilon})$. These conditions
define a diamond-shaped physical domain in the
$(\Delta\epsilon,\bar{\epsilon})$ plane, which provides a convenient
parameterization for analyzing the phase and bifurcation structure discussed
in the following subsections.

Using the activation probabilities $A_{n,d}^{-}(c)$ and
$A_{n,d}^{+}(c)$ defined in Sec.~\ref{sec:mean_field_model}, Eq.~\eqref{eq:model_drift} can be written as
\begin{equation}
v_{n,d}(c)
=
M_n(c)
+n\bar{\epsilon}
\left[
A_{n,d}^{-}(c)-A_{n,d}^{+}(c)
\right]
+n\Delta\epsilon
\left[
A_{n,d}^{-}(c)+A_{n,d}^{+}(c)
\right].
\label{eq:drift_ebar_deltae}
\end{equation}
This form separates the majority-rule contribution from the symmetric and asymmetric parts of collective reversal. The sign of $M_n(c)$ follows the population imbalance: it is positive above $c=1/2$ and negative below it. Majority rule, therefore, drives the population away from the mixed state. The term containing $\bar{\epsilon}$ has the opposite effect. For $c>1/2$, one has
$A_{n,d}^{+}(c)>A_{n,d}^{-}(c)$, making this contribution negative; the sign is reversed for $c<1/2$. Consequently, symmetric reversal acts as a restoring mechanism around $c=1/2$.

The contribution containing $\Delta\epsilon$ introduces the directional bias. Since $A_{n,d}^{-}(c)+A_{n,d}^{+}(c)>0$ for $0\leq c\leq1$, its direction is fixed by the sign of $\Delta\epsilon$. A positive value favors opinion $+1$, while a negative value favors opinion $-1$. Its magnitude varies with $c$ via the activation probabilities, so this bias is composition-dependent rather than uniform. The dissent tolerance determines which group compositions contribute to these reversal terms. When $d=0$, only the two unanimous configurations are eligible. For $d>0$, the eligible ranges become $\ell=0,1,\ldots,d$ and $\ell=n-d,\ldots,n$. The additional configurations are near-unanimous groups for which majority rule and collective reversal lead to opposite final states. At $c=1/2$, the probability of selecting a group from either eligible range is
\begin{equation}
\Phi_{n,d}
\equiv
A_{n,d}^{-}\left(\frac{1}{2}\right)
+
A_{n,d}^{+}\left(\frac{1}{2}\right)
=
2^{1-n}
\sum_{\ell=0}^{d}
\binom{n}{\ell}.
\label{eq:eligible_probability}
\end{equation}
The quantity $\Phi_{n,d}$ measures the availability of eligible groups. It should be distinguished from the probability of an actual reversal event, which is $\epsilon\Phi_{n,d}$ in the symmetric case. For strict unanimity,
$\Phi_{n,0}=2^{1-n}$. Allowing one dissenter increases this value to
$\Phi_{n,1}=2^{1-n}(n+1)$. Hence, changing $d$ leaves the conditional reversal probability $\epsilon$ unchanged but alters how often the reversal channel is available.

The exchange properties of the drift follow from
$M_n(1-c)=-M_n(c)$ and
$A_{n,d}^{+}(c)=A_{n,d}^{-}(1-c)$. The difference
$A_{n,d}^{-}-A_{n,d}^{+}$ is antisymmetric about $c=1/2$, while the sum
$A_{n,d}^{-}+A_{n,d}^{+}$ is symmetric. When
$\Delta\epsilon=0$, these relations give $v_{n,d}(1-c)=-v_{n,d}(c),$ and the mixed state $c^\ast=1/2$ is a fixed point. For
$\Delta\epsilon\neq0$, the exchange symmetry is broken, and the midpoint is generally shifted away from stationarity. A stationary state satisfies $v_{n,d}(c^\ast)=0$ and is linearly stable if
$v_{n,d}'(c^\ast)<0$. At the two boundaries, $v_{n,d}(0)=n\epsilon_{\uparrow}$ and $v_{n,d}(1)=-n\epsilon_{\downarrow}$ regardless of $d$. If both reversal probabilities are positive, the drift points into the physical interval at both ends, and at least one interior fixed point must exist. In the one-sided limits, one boundary remains absorbing: $c=1$ for
$\epsilon_{\downarrow}=0$ and $c=0$ for
$\epsilon_{\uparrow}=0$. The tolerance $d$ does not affect these boundary values but does change the number, positions, and stability of the interior fixed points.

\subsection{Linear stability and accessibility of the symmetric transition}
\label{sec:linear_stability}

The local stability of the mixed state can be determined by expanding the
macroscopic drift about $c=1/2$. Writing $c=1/2+\delta$, with
$|\delta|\ll1$, Eq.~\eqref{eq:drift_ebar_deltae} becomes
\begin{equation}
    v_{n,d}\left(\frac{1}{2}+\delta\right)
    =
    n2^{1-n}B_{n,d}\Delta\epsilon
    +
    \lambda_{n,d}(\bar{\epsilon})\delta
    +O(\delta^2),
    \label{eq:local_drift_compact_revised}
\end{equation}
where
$\lambda_{n,d}(\bar{\epsilon})
=\Lambda_n-n2^{2-n}S_{n,d}\bar{\epsilon}$,
$B_{n,d}=\sum_{\ell=0}^{d}\binom{n}{\ell}$, and
$S_{n,d}=\sum_{\ell=0}^{d}(n-2\ell)\binom{n}{\ell}$. The coefficient
associated with pure majority-rule ordering is (see Eq.~\eqref{eq:appendix_lambda_final} in Appendix~\ref{app_lambda} for details)
\begin{equation}
    \Lambda_n
    =
    2^{2-n}
    \sum_{k=1}^{d_{\max}}
    k(n-2k)\binom{n}{k},
    \label{eq:lambda_n}
\end{equation}
with $d_{\max} = \left\lfloor(n-1)/2\right\rfloor$. Because every
term contributing to $\Lambda_n$ is positive, pure majority rule renders the
mixed state linearly unstable: an infinitesimal population imbalance is
amplified rather than suppressed. Symmetric collective reversal counteracts
this ordering tendency through the contribution proportional to
$S_{n,d}\bar{\epsilon}$, whereas asymmetric reversal generates a finite drift
at the midpoint proportional to $B_{n,d}\Delta\epsilon$. For symmetric
reversal, $\Delta\epsilon=0$, the constant term vanishes, and the next
nonlinear correction is of order $O(\delta^3)$.

\begin{figure}[pos=tb!]
    \centering
    \includegraphics[width=\linewidth]{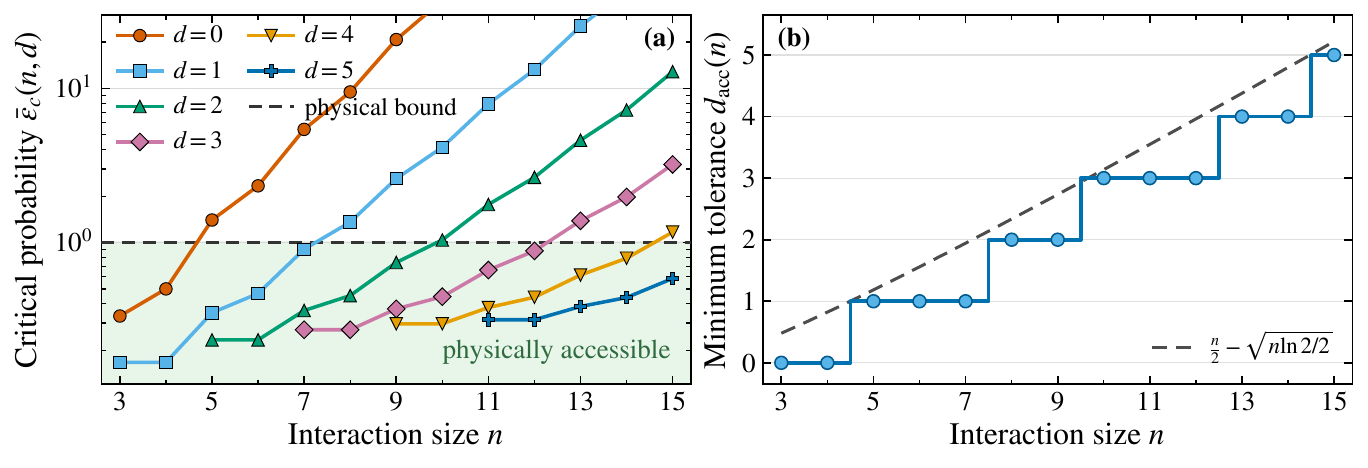}
    \caption{Accessibility of the symmetric mixed-state stability threshold.
    (a) Formal critical probability $\bar{\epsilon}_c(n,d)$ from
    Eq.~\eqref{eq:critical_ebar_revised} for different dissent tolerances.
    The horizontal dashed line marks the physical bound
    $\bar{\epsilon}=1$, and the shaded region contains accessible thresholds.
    (b) Exact minimum tolerance $d_{\mathrm{acc}}(n)$ from
    Eq.~\eqref{eq:d_accessibility}. The step line guides the eye, while the
    dashed curve shows the large-$n$ approximation in
    Eq.~\eqref{eq:d_accessibility_asymptotic}.}
    \label{fig:critical_probability}
\end{figure}

For symmetric reversal, $\Delta\epsilon=0$, the mixed state $c^\ast=1/2$
remains an exact fixed point. It is linearly unstable when
$\lambda_{n,d}(\bar{\epsilon})>0$ and stable when
$\lambda_{n,d}(\bar{\epsilon})<0$. Setting
$\lambda_{n,d}(\bar{\epsilon}_c)=0$ yields the critical reversal probability
for any $n\geq3$ and any admissible dissent tolerance
$0\leq d\leq d_{\max}$ (see Eq.~\eqref{eq:appendix_critical_compact} in Appendix~\ref{app_lambda} for details),
\begin{equation}
    \bar{\epsilon}_c(n,d)
    =
    \frac{
        n\binom{n-1}{d_{\max}}-2^{n-1}
    }{
        2n\binom{n-1}{d}
    }.
    \label{eq:critical_ebar_revised}
\end{equation}
The numerator in Eq.~\eqref{eq:critical_ebar_revised} quantifies the linear
ordering tendency generated by majority rule, whereas the denominator measures the restoring effect of collective reversal over all eligible group compositions. The critical probability is therefore controlled jointly by the
interaction size $n$ and the dissent tolerance $d$. Since
$0\leq\bar{\epsilon}\leq1$, stabilization of the mixed state is physically
accessible only when $\bar{\epsilon}_c(n,d)\leq1$. This accessibility
condition is illustrated in Fig.~\ref{fig:critical_probability}(a), which
shows $\bar{\epsilon}_c(n,d)$ for several fixed values of $d$. Under strict
unanimity, $d=0$, the thresholds
$\bar{\epsilon}_c(3,0)=1/3$ and $\bar{\epsilon}_c(4,0)=1/2$ lie within the
physical probability interval, whereas $\bar{\epsilon}_c(n,0)>1$ for every
$n\geq5$. The loss of accessibility at larger $n$ arises because unanimous
groups become increasingly rare near the mixed state, causing the reversal
channel to weaken relative to the ordering drift generated by the much more
frequent strict-majority configurations. Relaxing the activation condition to
$d>0$ includes near-unanimous groups, increases the statistical weight of the
eligible composition sectors, and consequently shifts the critical probability
downward. Physically accessible transitions can therefore reappear at larger
interaction sizes; for example,
$\bar{\epsilon}_c(5,1)=7/20$,
$\bar{\epsilon}_c(6,1)=7/15$, and
$\bar{\epsilon}_c(7,1)=19/21$, even though the corresponding
strict-unanimity thresholds lie outside the physical range.

This result also shows why the frequency of activation matters. If reversal requires complete unanimity, eligible groups are rarely encountered as $n$ increases, especially near $c=1/2$. A high reversal probability is insufficient to influence population-level dynamics. Setting $d>0$ relaxes this requirement by admitting groups with a small minority so that activation can occur without exact agreement. A similar threshold logic appears in models of collective behavior, where a response begins after social support reaches a prescribed level
\cite{granovetter1978threshold}. Studies of complex contagion likewise emphasize that adoption may depend on reinforcement from several contacts
\cite{guilbeault2021topological}, while experiments show that group pressure can affect individual judgments and political opinions
\cite{franzen2023power}. These studies motivate the use of a state-dependent activation rule but do not provide a direct calibration of $d$. In the present model, $d$ simply specifies the range of local compositions that can activate collective reversal.

To examine the effect of large interaction groups, we first keep $d$ fixed and take $n\rightarrow\infty$. Applying the central-binomial approximation to Eq.~\eqref{eq:critical_ebar_revised} gives [see Eq.~\eqref{eq:critical_leading} in Appendix~\ref{app_lambda}]
\begin{equation}
\bar{\epsilon}_c(n,d)
\sim
2^{n-2}d!\sqrt{\frac{2}{\pi}},
n^{-(d+1/2)}.
\label{eq:critical_fixed_d_asymptotic}
\end{equation}
For fixed $d$, the exponential factor $2^{n-2}$ eventually dominates the power-law factor $n^{-(d+1/2)}$, and $\bar{\epsilon}_c(n,d)$ increases rapidly with $n$. This behavior follows from the sampling statistics near the mixed state. The eligible groups contain no more than $d$ dissenters, so they remain confined to narrow composition ranges close to the two ordered limits. As $n$ grows, these ranges account for a progressively smaller fraction of all possible group compositions. Reversal events then become too rare to balance the majority-rule drift, even at the largest allowed value $\bar{\epsilon}=1$. Once $\bar{\epsilon}_c(n,d)>1$, the transition is no longer physically accessible.

A different behavior is obtained when $d$ increases with $n$. At
$d=d_{\max}=\lfloor(n-1)/2\rfloor$, every group with a strict local majority is eligible, including compositions close to an equal split. In this limit, $\bar{\epsilon}_c(n,d_{\max})<1/2$ for all $n$. Hence, a transition can remain accessible for large groups, but the number of tolerated dissenters cannot remain fixed.

The minimum tolerance needed to satisfy the physical bound can be defined as
\begin{equation}
    d_{\mathrm{acc}}(n)
    =
    \min\left\{
        d\in\mathbb{Z}:
        0\leq d\leq d_{\max},
        \ \bar{\epsilon}_c(n,d)\leq1
    \right\}.
    \label{eq:d_accessibility}
\end{equation}
The quantity $d_{\mathrm{acc}}(n)$ gives the smallest dissent tolerance for which collective reversal can stabilize the mixed state with
$0\leq\bar{\epsilon}\leq1$. Figure~\ref{fig:critical_probability}(b) shows that
$d_{\mathrm{acc}}(3)=d_{\mathrm{acc}}(4)=0$, so unanimity is sufficient for
$n=3$ and $n=4$. The minimum tolerance increases to
$d_{\mathrm{acc}}(5)=d_{\mathrm{acc}}(6)=d_{\mathrm{acc}}(7)=1$ and
$d_{\mathrm{acc}}(8)=d_{\mathrm{acc}}(9)=2$. For $n=5$, the eligible set must include the compositions $4{:}1$ and $1{:}4$ in addition to $5{:}0$ and $0{:}5$. For $n=8$, the minimum accessible choice $d=2$ further includes compositions down to $6{:}2$ and $2{:}6$. Because $d$ takes integer values, the same minimum tolerance can apply to several consecutive group sizes. This produces the staircase pattern in Fig.~\ref{fig:critical_probability}(b).

The large-$n$ form of this boundary is derived in Appendix~\ref{app_lambda} [see Eq.~\eqref{eq:critical_dcc}] and is given by
\begin{equation}
d_{\mathrm{acc}}(n)
\simeq
d_{\max}
-
\sqrt{\frac{(n-1)\ln 2}{2}}.
\label{eq:d_accessibility_asymptotic}
\end{equation}
Using $d_{\max}\simeq n/2$, this becomes
$d_{\mathrm{acc}}(n)\simeq n/2-\sqrt{n\ln2/2}$. The dashed curve in
Fig.~\ref{fig:critical_probability}(b) shows this asymptotic estimate. It is included as a large-$n$ reference, not as a fit to the discrete data. The exact boundary changes in integer steps, while Eq.~\eqref{eq:d_accessibility_asymptotic} is continuous, so differences are expected for the relatively small values of $n$ displayed.

The dominant contribution to $d_{\mathrm{acc}}$ is proportional to $n$, with a correction of order $\sqrt{n}$. The corresponding minimum fraction of aligned agents is $1-d_{\mathrm{acc}}/n\simeq1/2+\sqrt{\ln2/(2n)}$ which approaches $1/2$ from above. In large groups, an accessible reversal mechanism must therefore include groups with a majority only slightly larger than an equal split. Restricting activation to near-unanimous groups would make reversal too infrequent. In a social interpretation, a counternarrative or institutional response based on an almost unanimous local consensus would face the same sampling limitation as the group size increases. This comparison concerns only the model's activation statistics and does not provide an empirical threshold for real interventions.

\subsection{Phase diagram and asymmetric bifurcations}
\label{sec:phase_diagram}

For fixed $n$ and $d$, the stationary macroscopic states are given by the
physical roots of $v_{n,d}(c^\ast)=0$, with stability determined by
$v_{n,d}'(c^\ast)<0$. When both reversal probabilities are positive,
$v_{n,d}(0)>0$ and $v_{n,d}(1)<0$, ensuring the existence of at least one
interior fixed point. In the parameter ranges considered here, the dynamics is
either monostable, with a single stable interior fixed point, or bistable, with
two stable fixed points separated by an unstable one. The boundary between
these regimes occurs when a stable and an unstable fixed point coalesce in a
saddle-node bifurcation, $ v_{n,d}(c_{\mathrm{sn}})=0, \, v_{n,d}'(c_{\mathrm{sn}})=0,$ where $0<c_{\mathrm{sn}}<1$.

To express the saddle-node boundary in the
$(\Delta\epsilon,\bar{\epsilon})$ plane, we define
$D_{n,d}(c)=A_{n,d}^{-}(c)-A_{n,d}^{+}(c)$ and
$Q_{n,d}(c)=A_{n,d}^{-}(c)+A_{n,d}^{+}(c)$. Since
$v_{n,d}=M_n+n\bar{\epsilon}D_{n,d}
+n\Delta\epsilon Q_{n,d}$ is linear in both reversal parameters, the two
conditions, namely $v_{n,d}(c_{\mathrm{sn}})=0, \, v_{n,d}'(c_{\mathrm{sn}})=0$, give the parametric
boundary
    \begin{align}
    \bar{\epsilon}_{\mathrm{sn}}(c)
    &=
    \frac{
        Q_{n,d}(c)M_n'(c)-M_n(c)Q_{n,d}'(c)
    }{
        n\mathcal{J}_{n,d}(c)
    }, \label{eq:saddle_node_parametric_1}\\
    \Delta\epsilon_{\mathrm{sn}}(c)
    &=
    \frac{
        M_n(c)D_{n,d}'(c)-D_{n,d}(c)M_n'(c)
    }{
        n\mathcal{J}_{n,d}(c)
    },
    \label{eq:saddle_node_parametric}        
    \end{align}
Here,
$\mathcal{J}_{n,d}(c)=D_{n,d}(c)Q_{n,d}'(c)
-D_{n,d}'(c)Q_{n,d}(c)$, with
$\mathcal{J}_{n,d}(c)\neq0$ along the regular parametric boundary.
Further details are provided in Appendix~\ref{app_saddle_node}. Only curve
segments satisfying $0\leq\bar{\epsilon}\leq1$ and
$|\Delta\epsilon|\leq\min(\bar{\epsilon},1-\bar{\epsilon})$ belong to the
physical probability domain. The phase diagram is symmetric about
$\Delta\epsilon=0$ because
$v_{n,d}(1-c;\bar{\epsilon},\Delta\epsilon)
=-v_{n,d}(c;\bar{\epsilon},-\Delta\epsilon)$.

Along the symmetric line $\Delta\epsilon=0$, the mixed state $c^\ast=1/2$ is
always stationary. When the threshold $\bar{\epsilon}_c(n,d)$ lies within the
physical probability domain, the mixed state changes stability at the value
given by Eq.~\eqref{eq:critical_ebar_revised}. For the parameter combinations
examined below, this transition is a supercritical pitchfork bifurcation:
below $\bar{\epsilon}_c$, the mixed state is unstable and separates two
symmetry-related stable ordered states; at $\bar{\epsilon}_c$, the ordered
branches merge continuously with the mixed state, and above
$\bar{\epsilon}_c$, the mixed state is the unique stable solution. Locally,
the symmetric drift has the normal form $v_{n,d}
    =
    \lambda_{n,d}(\bar{\epsilon})\delta
    -
    g_{n,d}(\bar{\epsilon})\delta^3
    +
    O(\delta^5).$
At the corresponding critical points, direct expansion gives
$g_{5,1}(\bar{\epsilon}_c)=25$ and
$g_{8,2}(\bar{\epsilon}_c)=248/3$, confirming that both coefficients are
positive. Consequently, the ordered solutions satisfy
$|\delta^\ast|\sim
[\lambda_{n,d}(\bar{\epsilon})/g_{n,d}(\bar{\epsilon})]^{1/2}
\propto(\bar{\epsilon}_c-\bar{\epsilon})^{1/2}$ as the transition is
approached from the ordered side.

\begin{figure}[pos=tb!]
\centering
\includegraphics[width=\linewidth]{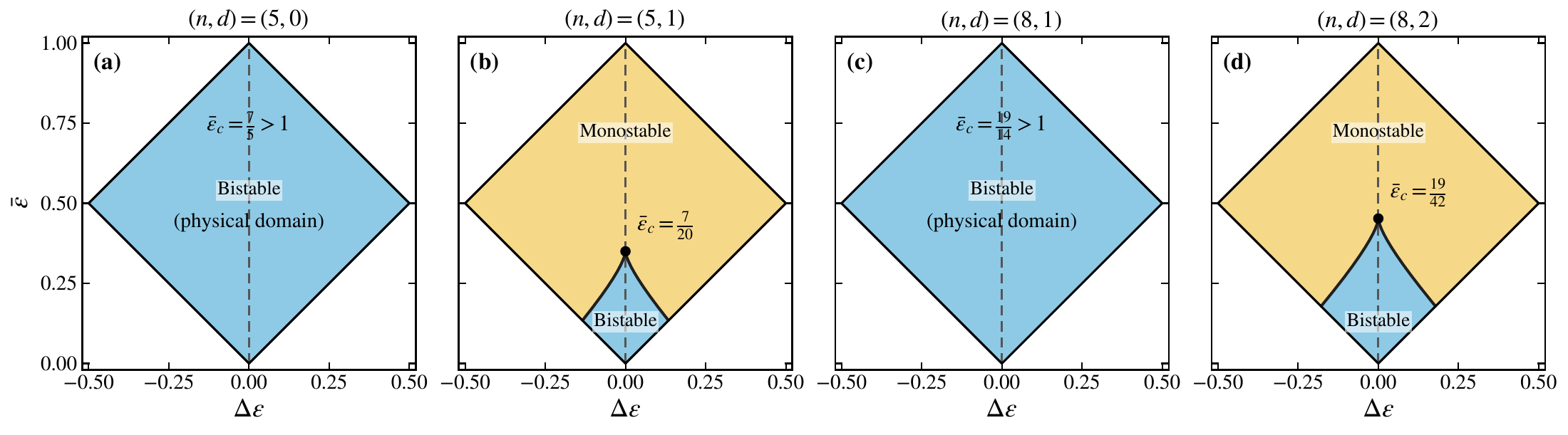}
\caption{Mean-field phase diagrams in the
$(\Delta\epsilon,\bar{\epsilon})$ plane for
(a) $(n,d)=(5,0)$, (b) $(5,1)$,
(c) $(8,1)$, and (d) $(8,2)$. Blue regions are bistable, and yellow regions are monostable. Solid black curves mark the double-root boundaries obtained from
Eqs.~\eqref{eq:saddle_node_parametric_1} and
\eqref{eq:saddle_node_parametric}. For
$\Delta\epsilon\neq0$, these curves correspond to saddle-node bifurcations. The vertical dashed lines show the symmetric case
$\Delta\epsilon=0$, and the black circles mark the accessible pitchfork thresholds from Eq.~\eqref{eq:critical_ebar_revised}. The diamond-shaped outline is the physical domain
$0\leq\epsilon_{\uparrow},\epsilon_{\downarrow}\leq1$.}
\label{fig:phase_diagram}
\end{figure}

Figure~\ref{fig:phase_diagram} compares four combinations of $n$ and $d$. For $(n,d)=(5,0)$, the symmetric threshold
$\bar{\epsilon}_c=7/5$ lies above the physical bound. No monostable region is then found within the probability domain, as shown in Fig.~\ref{fig:phase_diagram}(a). With one tolerated dissenter, the threshold decreases to
$\bar{\epsilon}_c(5,1)=7/20$. The pitchfork point and its two adjoining saddle-node curves now lie inside the physical domain, giving the monostable region in Fig.~\ref{fig:phase_diagram}(b). The results for $n=8$ follow the same pattern. At $d=1$, the threshold $\bar{\epsilon}_c(8,1)=19/14$ is inaccessible, and the physical domain remains bistable [Fig.~\ref{fig:phase_diagram}(c)]. Increasing the tolerance to $d=2$ lowers the threshold to $\bar{\epsilon}_c(8,2)=19/42$, producing the monostable region in Fig.~\ref{fig:phase_diagram}(d). These examples illustrate the different roles of $n$ and $d$. Increasing $n$ reduces the probability of sampling a highly homogeneous group near $c=1/2$. Increasing $d$ increases the probability of group compositions in the reversal channel and can bring the transition back into the physical range.

Away from the symmetric line, the pitchfork separates into two saddle-node boundaries. Consider $\Delta\epsilon>0$, for which reversal favors opinion $+1$. The unstable fixed point separating the two basins shifts toward the negative-majority branch, and the basin of the positive-majority attractor expands. At the saddle-node boundary, the negative-majority attractor merges with the unstable fixed point and disappears. Only the positive-majority attractor remains. The corresponding process for $\Delta\epsilon<0$ is obtained by exchanging the two opinions. Before either boundary is crossed, asymmetry changes the stationary states and their basin sizes but does not remove bistability.

The distinction between the two phase regions is dynamical. In the bistable region, the long-time state depends on the initial concentration because the two stable fixed points have different basins of attraction. In the monostable region, every interior initial condition approaches the same stable branch. Changing $d$ is not equivalent to changing
$\epsilon_{\uparrow}$ or $\epsilon_{\downarrow}$. The tolerance modifies the set of group compositions that are exposed to reversal, whereas the two probabilities determine a group's response once it becomes eligible. A broader set of eligible options can therefore remove a competing attractor even when the conditional reversal probabilities remain fixed.

These phase diagrams describe deterministic MF dynamics. In a finite population, fluctuations can shift the observed tipping point and may induce rare transitions between coexisting attractors. It is also important to distinguish the local and global conditions used in the analysis. The inequality
$\bar{\epsilon}_c(n,d)\leq1$ determines whether stabilization of the mixed state is accessible along $\Delta\epsilon=0$. It does not determine the number of fixed points over the entire asymmetric domain. The fully bistable regions in
Figs.~\ref{fig:phase_diagram}(a) and \ref{fig:phase_diagram}(c) are obtained from the roots and stability of the complete drift, rather than from the symmetric linear-stability criterion alone.

\subsection{Stationary branches and macroscopic symmetry restoration}
\label{sec:order_disorder}

The phase diagrams in Sec.~\ref{sec:phase_diagram} identify the number and
stability of the macroscopic stationary states. To determine how these states
move as the reversal probability changes, we now express the stationary
solutions in terms of the order parameter (magnetization) $m=2c-1$. The mixed state corresponds
to $m^\ast=0$, whereas $m^\ast\neq0$ indicates persistent macroscopic dominance of one opinion. Along the symmetric line
$\epsilon_{\uparrow}=\epsilon_{\downarrow}=\epsilon$, the drift becomes
\begin{equation}
    v_{n,d}(c)
    =
    M_n(c)
    +
    n\epsilon
    \left[
        A_{n,d}^{-}(c)-A_{n,d}^{+}(c)
    \right].
    \label{eq:drift_symmetric_c}
\end{equation}
Under the transformation $c=(1+m)/2$, the exchange symmetry
$c\leftrightarrow1-c$ becomes $m\leftrightarrow-m$. Because
$v_{n,d}(c)=dc/d\tau$, the corresponding magnetization drift is defined as $\widetilde v_{n,d}(m;\epsilon)\equiv dm/d\tau = 2v_{n,d}\!\left((1+m)/2;\epsilon\right)$. For symmetric reversal, $\widetilde v_{n,d}(m;\epsilon)$ is an odd function
of $m$ and can therefore be written as
$\widetilde v_{n,d}(m;\epsilon)
=m\mathcal{P}_{n,d}(m^2;\epsilon)$. The mixed state $m^\ast=0$ is thus always
a stationary solution. The nonzero stationary branches are obtained from the
physical positive roots $y^\ast=m^{\ast2}$ of
$\mathcal{P}_{n,d}(y^\ast;\epsilon)=0$, with
$0<y^\ast\leq1$, giving
$m_\pm^\ast=\pm\sqrt{y^\ast}$.

The effect of relaxing the activation condition can be shown explicitly for
$n=5$. Under strict unanimity, $d=0$, the physical ordered branches are
\begin{equation}
    m_\pm^\ast(\epsilon)
    =
    \pm
    \left[
        \frac{
            5(1+\epsilon)
            -
            2\sqrt{1+18\epsilon+5\epsilon^2}
        }{
            3-\epsilon
        }
    \right]^{1/2}.
    \label{eq:m5_d0}
\end{equation}
Their formal merging point,
$\epsilon_c(5,0)=7/5$, lies outside the admissible probability interval.
Accordingly, the branches approach the mixed state as $\epsilon$ increases but remain separated from it throughout $0\leq\epsilon\leq1$. Allowing one dissenter, $d=1$, changes the physical branches to
\begin{equation}
    m_\pm^\ast(\epsilon)
    =
    \pm
    \left[
        \frac{
            5
            -
            2\sqrt{1+8\epsilon+20\epsilon^2}
        }{
            3+4\epsilon
        }
    \right]^{1/2},
    \label{eq:m5_d1}
\end{equation}
where $ 0\leq\epsilon\leq 7/20$. These branches now merge continuously with $m^\ast=0$ at the physically
accessible threshold $\epsilon_c(5,1)=7/20$. The algebra leading to
Eqs.~\eqref{eq:m5_d0} and \eqref{eq:m5_d1} is provided in
Appendix~\ref{app_stationary_branches} [see Eqs.~\eqref{eq:mag_app_1} and \eqref{eq:mag_app_2}]. The comparison shows that increasing
$d$ does more than reduce the magnitude of the stationary ordering: it can
bring the branch-merging point into the physical domain and thereby permit
macroscopic symmetry restoration.

Near any accessible symmetric threshold considered here, the stationary
ordered branches follow $|m^\ast| \sim (\epsilon_c(n,d)-\epsilon)^{1/2}$ as $\epsilon\rightarrow\epsilon_c^{-}$. The MF order-parameter exponent is therefore $\beta=1/2$, consistent with a supercritical pitchfork bifurcation. Below the threshold, the two opinions are microscopically equivalent, but the stationary population
spontaneously selects one ordered branch. At the threshold, both branches
merge continuously with the mixed state, which becomes stable above
$\epsilon_c$. This behavior is not restricted to $n=5$. For $n=8$, the
threshold $\epsilon_c(8,1)=19/14$ is inaccessible, and the ordered branch
survives over the complete physical interval, whereas increasing the tolerance
to $d=2$ lowers the threshold to $\epsilon_c(8,2)=19/42$ and restores a
continuous transition.

\begin{figure}[pos=tb!]
\centering
\includegraphics[width=0.9\linewidth]{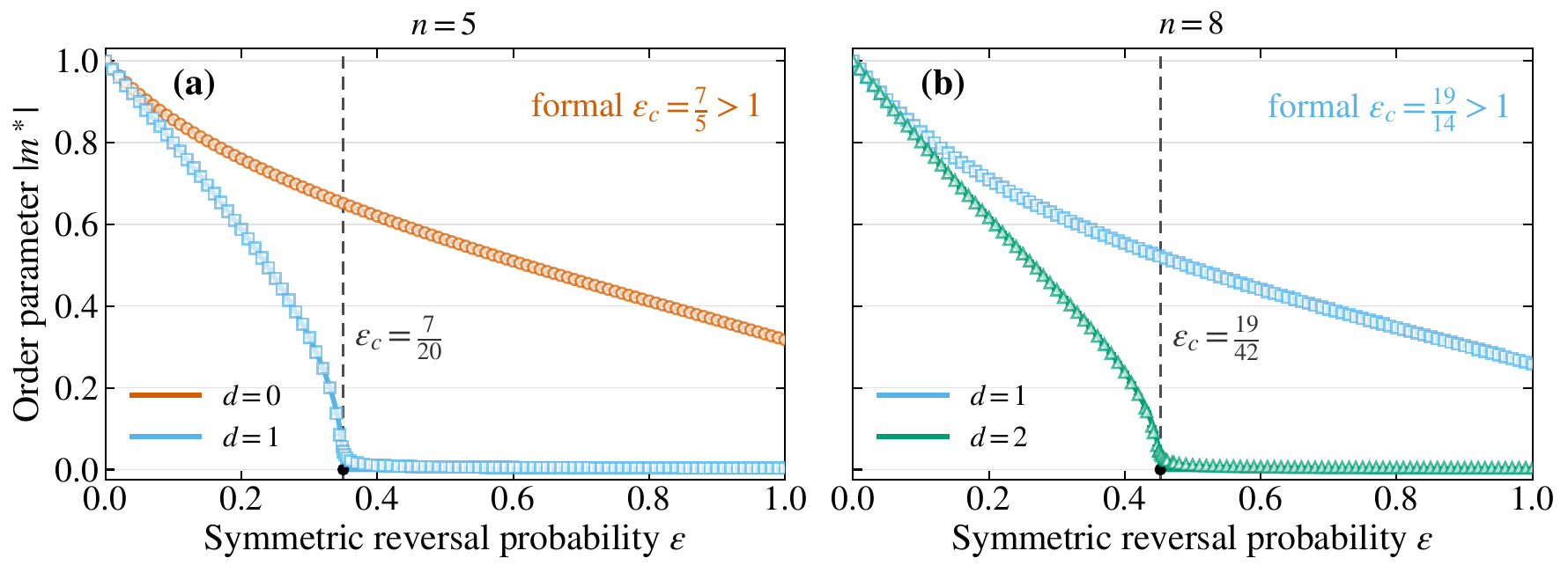}
\caption{Stationary absolute order parameter as a function of the symmetric reversal probability $\epsilon$ for
(a) $n=5$ with $d=0$ and $1$, and
(b) $n=8$ with $d=1$ and $2$. Solid curves are the stable roots of
Eq.~\eqref{eq:drift_symmetric_c}, and open symbols show MC results for
$N=10^5$. Vertical dashed lines and black circles indicate the accessible thresholds
$\epsilon_c(5,1)=7/20$ and
$\epsilon_c(8,2)=19/42$. Error bars are standard errors and are omitted when smaller than the symbols.}
\label{fig:mag}
\end{figure}

Figure~\ref{fig:mag} compares the stationary MF solutions with the MC measurements. For $(n,d)=(5,0)$ and $(8,1)$, the formal thresholds lie above $\epsilon=1$. The order parameter decreases as $\epsilon$ increases but remains nonzero over the physical interval. A different behavior appears for $(n,d)=(5,1)$ and $(8,2)$, where the ordered branches reach zero continuously at the predicted thresholds. The MC data agree with the analytical results, apart from rounding near the transition due to the finite population size. The state $m^\ast=0$ refers to equal macroscopic fractions of the two opinions. It does not mean that individual agents become neutral, since every agent remains in either state $+1$ or $-1$. In a finite system, fluctuations around equal fractions give a small positive value of $\langle |m|\rangle$ and smooth the transition near $\epsilon_c$. This residual contribution decreases with $N$, bringing the simulation results closer to the MF branches.

\begin{figure}[pos=tb!]
\centering
\includegraphics[width=\linewidth]{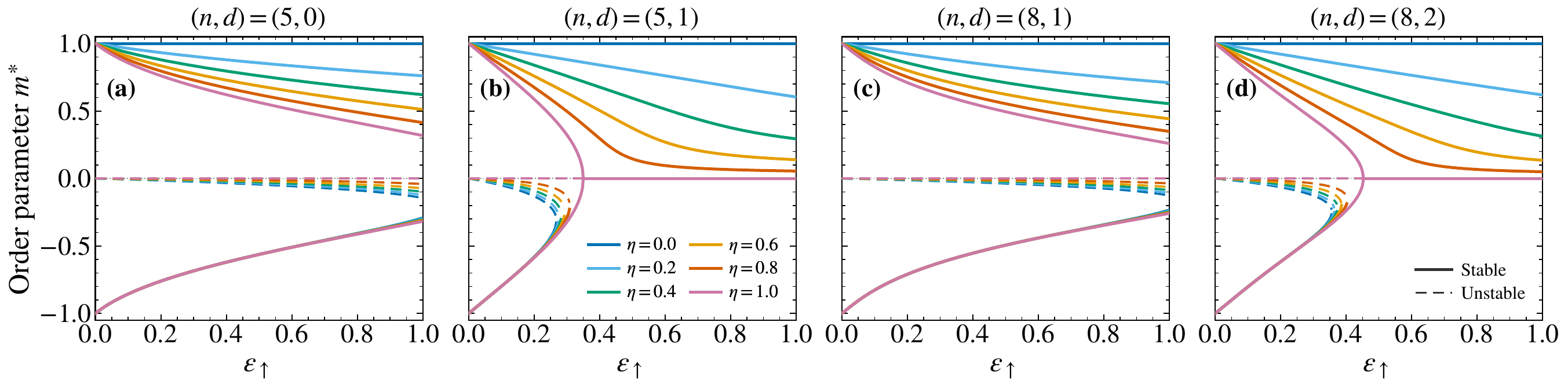}
\caption{Mean-field stationary branches satisfying
$v_{n,d}(c^\ast)=0$ along
$\epsilon_{\downarrow}=\eta\epsilon_{\uparrow}$ for
$\eta=0,0.2,\ldots,1$: (a) $(n,d)=(5,0)$,
(b) $(5,1)$, (c) $(8,1)$, and (d) $(8,2)$.
Solid curves denote stable fixed points, and dashed curves denote unstable fixed points. Panels (a) and (c) remain bistable over the physical range. In panels (b) and (d), bistability ends at a pitchfork bifurcation for $\eta=1$ and at a saddle-node bifurcation for $\eta<1$.}
\label{fig:mag_asym}
\end{figure}

We now consider unequal reversal probabilities along $\epsilon_{\downarrow}=\eta\epsilon_{\uparrow}$, with $0\leq\eta\leq1$. 
The endpoints $\eta=1$ and $\eta=0$ correspond to symmetric and one-sided reversal, respectively. Along each path, $\bar{\epsilon}=(1+\eta)\epsilon_{\uparrow}/2$ and
$\Delta\epsilon=(1-\eta)\epsilon_{\uparrow}/2$. For $\eta<1$, the reversal channel is biased toward opinion $+1$. Figure~\ref{fig:mag_asym} shows how this bias changes the stationary branches. For $(n,d)=(5,0)$ and $(8,1)$, the activation range remains too narrow to remove either ordered attractor within the physical probability interval. The locations of the two stable branches and the unstable fixed point change with $\epsilon_{\uparrow}$, but bistability persists. For $(n,d)=(5,1)$ and $(8,2)$, a sufficiently broad activation range allows one branch to disappear. At $\eta=1$, the two ordered branches meet the mixed branch at the pitchfork bifurcation. For each asymmetric path $\eta<1$ shown in the figure, the negative-majority branch merges with the intermediate unstable branch at a saddle-node point. After this collision, only the positive-majority branch remains stable.

Directional bias and the loss of bistability are separate effects. Before the saddle-node point, the basin of the positive-majority attractor expands, but initial conditions inside the negative basin can still approach the negative-majority branch. Beyond the saddle-node point, that branch no longer exists, and all interior trajectories converge to the favored state. The role of $d$ is to determine whether this collision can occur within the physical probability range. Increasing $d$ does not change the reversal probability of an eligible group; it increases the number and statistical weight of the compositions exposed to reversal. The condition $\eta<1$ represents an influence favoring one opinion. Before the saddle-node bifurcation, this influence biases the macroscopic outcome but does not remove its dependence on the initial state. After the bifurcation, the MF dynamics has a single stable outcome. This statement applies to the deterministic well-mixed model; finite populations, heterogeneous interactions, and time-dependent influences may behave differently.

\subsection{Consensus time}
\label{sec:consensus_time}

When both reversal probabilities are positive, neither $c=0$ nor $c=1$ is
absorbing: after reaching either boundary, the population can subsequently
leave it through collective reversal. We therefore define an absorbing-consensus time only in the one-sided limits, where one consensus configuration remains
absorbing. For $\epsilon_{\downarrow}=0$, the all-$+1$ state is absorbing,
whereas for $\epsilon_{\uparrow}=0$, the all-$-1$ state is absorbing. A
deterministic trajectory evolving monotonically from $c_0$ to $c_f$ requires
the time
\begin{equation}
    \tau(c_0\rightarrow c_f)
    =
    \int_{c_0}^{c_f}
    \frac{dc}{v_{n,d}(c)},
    \label{eq:tau_general_integral}
\end{equation}
provided that the drift does not vanish or change sign within the integration
interval. Since the drift vanishes at an absorbing boundary, the continuum trajectory reaches exact consensus only asymptotically. For a finite
population, we terminate the deterministic trajectory at $c_f=1-1/N$ or
$c_f=1/N$, corresponding to one remaining minority agent
~\cite{van1992stochastic,muslim2025ordering,muslim2026random}. The resulting integral is a
finite-size deterministic estimate, while the MC quantity is the
stochastic first-passage time to the fully absorbing configuration.

Consider first $\epsilon_{\downarrow}=0$, for which the drift toward positive
consensus is
$v_{n,d}^{(+)}(c)=M_n(c)+n\epsilon_{\uparrow}A_{n,d}^{-}(c)$. If $c_0$
belongs to the basin of $c=1$, the corresponding deterministic estimate is
\begin{equation}
    \tau_{n,d}^{(+)}
    (c_0;N,\epsilon_{\uparrow})
    =
    \int_{c_0}^{1-1/N}
    \frac{dc}{
        M_n(c)+n\epsilon_{\uparrow}A_{n,d}^{-}(c)
    }.
    \label{eq:tau_plus_general}
\end{equation}
If an unstable interior fixed point separates two deterministic basins,
Eq.~\eqref{eq:tau_plus_general} applies only to initial conditions on the
side flowing toward $c=1$. Numerical evaluation of this full integral provides the deterministic prediction used below for comparison with simulations.

The leading system-size dependence follows from the drift near the absorbing
boundary. Let $x=1-c$ denote the remaining fraction of agents holding opinion
$-1$. For $x\ll1$, the majority-rule contribution satisfies
$M_n(1-x)=nx+O(x^2)$, while the probability of selecting an eligible
negative-majority group is
$A_{n,d}^{-}(1-x)=\binom{n}{d}x^{n-d}
+O(x^{n-d+1})$. Hence,
\begin{equation}
    v_{n,d}^{(+)}(1-x)
    =
    nx
    +
    n\epsilon_{\uparrow}\binom{n}{d}x^{n-d}
    +
    O(x^2)
    +
    O\!\left(x^{n-d+1}\right).
    \label{eq:plus_boundary_drift}
\end{equation}
The term proportional to $x^{n-d}$ is the leading boundary contribution from
collective reversal. Since every admissible tolerance satisfies
$d\leq d_{\max}=\lfloor(n-1)/2\rfloor$, one always has $n-d>1$. The reversal
term is therefore asymptotically subleading relative to the linear
majority-rule contribution $nx$, even when the activation condition is
relaxed to its maximum admissible width.

Retaining the linear majority contribution and the leading reversal
contribution in Eq.~\eqref{eq:plus_boundary_drift}, and defining
$p=n-d-1$ and
$a_{n,d}^{(+)}=\epsilon_{\uparrow}\binom{n}{d}$, gives the boundary estimate as
\begin{equation}
    \tau_{n,d}^{(+)}
    \simeq
    \frac{1}{n}
    \left[
        \ln\!\bigl(N(1-c_0)\bigr)
        -
        \frac{1}{p}
        \ln\!\left(
            \frac{
                1+a_{n,d}^{(+)}(1-c_0)^p
            }{
                1+a_{n,d}^{(+)}N^{-p}
            }
        \right)
    \right].
    \label{eq:tau_plus_boundary}
\end{equation}
The derivation of Eq.~\eqref{eq:tau_plus_boundary} is provided in
Appendix~\ref{app_consensus_time}. Equation~\eqref{eq:tau_plus_boundary} is
not intended as a systematic next-to-leading expansion of the full drift; rather, it retains the leading reversal-dependent boundary contribution in
closed form. Higher-order majority-rule terms may modify the additive
constant but do not change the coefficient of $\ln N$. Because $N^{-p}\rightarrow0$ as $N\rightarrow\infty$, the leading finite-size
behavior is
\begin{equation}
    \tau_{n,d}^{(+)}
    =
    \frac{1}{n}\ln N
    +
    C_{n,d}^{(+)}(c_0,\epsilon_{\uparrow})
    +
    o(1),
    \label{eq:tau_plus_scaling}
\end{equation}
where $C_{n,d}^{(+)}$ is independent of $N$. Thus, broadening the activation window changes the finite and transient contributions to the consensus time without modifying its leading logarithmic scaling.

The logarithmic dependence comes from the last stage of the dynamics, when only a small negative minority remains. Let $x=1-c$ denote its population fraction. Near $x=0$, the probability of selecting a group containing a minority agent is proportional to $x$. Majority rule removes such agents at a rate
$v_{n,d}^{(+)}(1-x)\simeq nx$. The time spent in this boundary region is then on a logarithmic scale as $n^{-1} \ln N$, in agreement with the consensus kinetics of the well-mixed majority-rule model
\cite{krapivsky2003dynamics}. The reversal term has a higher-order dependence on the minority fraction. An eligible negative-majority group must contain at least $n-d$ negative agents, and its sampling probability scales as $x^{n-d}$. Since $n-d>1$, this contribution vanishes faster than $nx$ as the absorbing boundary is approached. Collective reversal can shorten the earlier part of the trajectory and enlarge the basin leading to $c=1$, but it has little effect on the final removal of the minority.

For a trajectory whose drift remains positive between $c_0$ and $1$, increasing $\epsilon_{\uparrow}$ or $d$ reduces the deterministic consensus time. The parameter $\epsilon_{\uparrow}$ increases the reversal contribution for each eligible group. Increasing $d$ adds further compositions to $A_{n,d}^{-}(c)$, making the same channel active more often. Both changes affect the finite term in Eq.~\eqref{eq:tau_plus_scaling}. The coefficient of $\ln N$ remains $1/n$ because it is fixed by the linear majority-rule drift near $c=1$.

\begin{figure}[pos=tb!]
\centering
\includegraphics[width=0.9\linewidth]{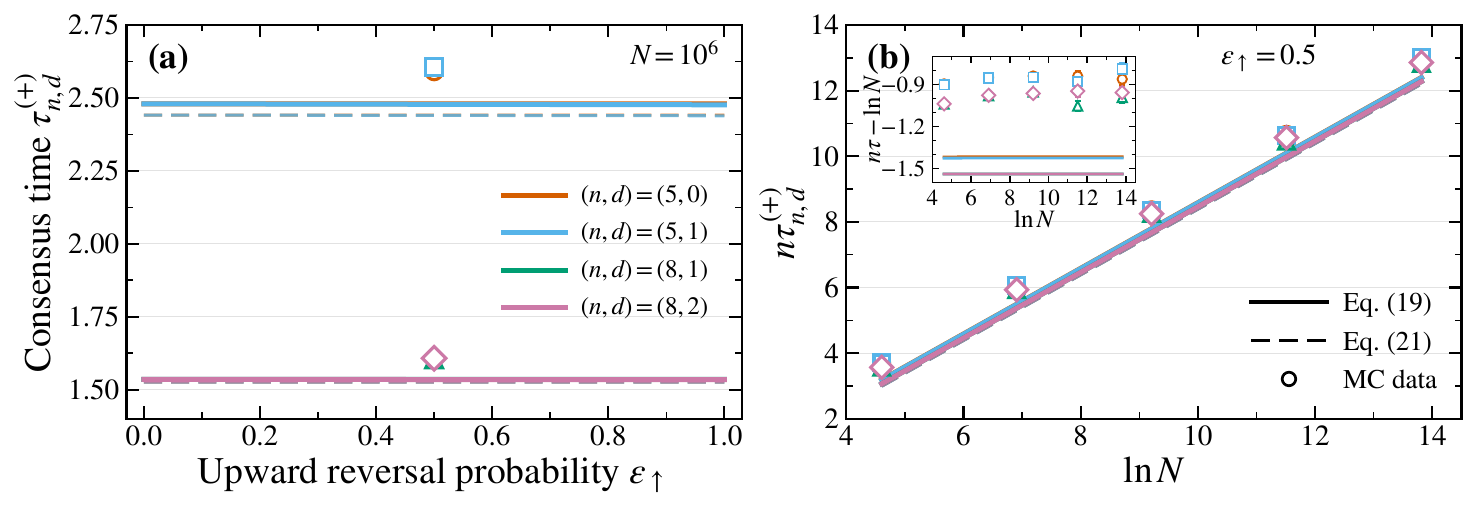}
\caption{Consensus time to the absorbing all-$+1$ state under one-sided reversal, $\epsilon_{\downarrow}=0$, for
$(n,d)=(5,0)$, $(5,1)$, $(8,1)$, and $(8,2)$.
(a) $\tau_{n,d}^{(+)}$ as a function of $\epsilon_{\uparrow}$ at
$N=10^6$; MC symbols are shown at $\epsilon_{\uparrow}=0.5$.
(b) Rescaled consensus time $n\tau_{n,d}^{(+)}$ versus $\ln N$ at
$\epsilon_{\uparrow}=0.5$. The inset shows
$n\tau_{n,d}^{(+)}-\ln N$. Solid curves are obtained from the full deterministic integral in Eq.~\eqref{eq:tau_plus_general}, dashed curves show the boundary approximation in
Eq.~\eqref{eq:tau_plus_boundary}, and symbols are MC first-passage results. Simulations start from $c_0=0.8$. Error bars are standard errors and are omitted when smaller than the symbols.}
\label{fig:consensus_time}
\end{figure}

Figure~\ref{fig:consensus_time}(a) shows the dependence on
$\epsilon_{\uparrow}$ for the four representative $(n,d)$ pairs. At fixed $n$, the larger tolerance gives a shorter consensus time because more negative-majority compositions are eligible for upward reversal. Panel (b) plots
$n\tau_{n,d}^{(+)}$ against $\ln N$. According to
Eq.~\eqref{eq:tau_plus_scaling}, all curves should approach a unit slope. The effect of $d$ and $\epsilon_{\uparrow}$ then appears mainly through their vertical offsets. The deterministic and MC results have nearly the same large-$N$ slope, although the MC values lie systematically higher. The deterministic integral stops at $c=1-1/N$, when one negative agent remains. The MC first-passage time also includes the extinction of this last agent and the fluctuations preceding absorption. These contributions produce an additional $O(1)$ time but do not change the coefficient of $\ln N$. This is reflected in the approximately parallel curves in Fig.~\ref{fig:consensus_time}(b).

The result for consensus toward opinion $-1$ follows from exchange symmetry. If $\epsilon_{\uparrow}=0$, the all-$-1$ state is absorbing and $v_{n,d}^{(-)}(c)
=
M_n(c)-n\epsilon_{\downarrow}A_{n,d}^{+}(c).$ The two consensus directions satisfy $\tau_{n,d}^{(-)}(c_0;N,\epsilon)
=
\tau_{n,d}^{(+)}(1-c_0;N,\epsilon),$
and therefore $\tau_{n,d}^{(-)}
=
n^{-1}\ln N
+
C_{n,d}^{(-)}
+
o(1).$ The leading coefficient is the same in both directions, so a separate numerical calculation for negative consensus would give a symmetry-related result.

The role of $d$ in the consensus time differs from its role in the stationary phase diagram. Near $c=1/2$, changing $d$ can shift the stability threshold into the physical range and remove one of the ordered attractors. Close to an absorbing boundary, however, the linear majority-rule term dominates regardless of $d$. One-sided reversal can steer the early evolution toward a preferred opinion, whereas ordinary majority updates control the final elimination of a dilute minority. If both $\epsilon_{\uparrow}$ and $\epsilon_{\downarrow}$ are positive, neither consensus boundary is absorbing. The appropriate quantity is then the relaxation time toward an interior stable fixed point. Locally, this time is proportional to
$|v_{n,d}'(c^\ast)|^{-1}$ and increases near a pitchfork or saddle-node bifurcation as the restoring drift becomes weak.

\section{Summary and conclusions}
\label{sec:summary}

We studied majority-rule dynamics with collective reversal in a well-mixed population. Reversal is restricted to groups containing at most $d$ dissenters, with $d=0$ corresponding to strict unanimity. The probabilities $\epsilon_{\uparrow}$ and $\epsilon_{\downarrow}$ determine the response of an eligible group, while $d$ determines which group compositions are eligible. This separation between response probability and activation range is the main extension of the unanimity-based model. The MF analysis gives a critical probability $\bar{\epsilon}_c(n,d)$ for the stability of the mixed state. Under strict unanimity, the threshold lies within the physical interval only for $n=3$ and $n=4$, where $\bar{\epsilon}_c=1/3$ and $1/2$, respectively. For $n\geq5$, unanimous groups occur too rarely near $c=1/2$ to balance the ordering drift. Allowing a small number of dissenters changes this result. For example,
$\bar{\epsilon}_c(5,1)=7/20$ and $\bar{\epsilon}_c(8,2)=19/42$, so transitions that are inaccessible under more restrictive activation become physically possible.

The minimum tolerance required for an accessible transition defines a boundary in the $(n,d)$ plane. If $d$ is held fixed, the eligible composition ranges become exponentially rare as $n$ increases, and $\bar{\epsilon}_c(n,d)$  eventually exceeds unity. Large groups, therefore, require a tolerance that grows with $n$. Its asymptotic form is $d_{\mathrm{acc}}(n)
\simeq
n/2
-
\sqrt{n\ln 2/2}$,
up to integer and subleading corrections. The associated majority fraction approaches $1/2$ from above. At the largest admissible tolerance,
$d=d_{\max}=\lfloor(n-1)/2\rfloor$, every strict-majority configuration is eligible, and the transition remains accessible for arbitrary $n$. The fixed cutoff between $n=4$ and $n=5$ obtained under unanimity is replaced by an accessibility boundary that depends on both $n$ and $d$.

For the accessible symmetric cases studied here, the ordered branches meet the mixed state through a supercritical pitchfork bifurcation, with
$|m^\ast|\propto(\bar{\epsilon}_c-\bar{\epsilon})^{1/2}$. The MC order parameter follows the stationary MF branches and reproduces the predicted thresholds, apart from finite-size rounding near criticality. Unequal reversal probabilities break the opinion-exchange symmetry. When the activation range is sufficiently broad, the disfavored stable branch collides with the intermediate unstable branch at a saddle-node bifurcation and disappears. The dynamics then changes from bistable to monostable. For the more restrictive cases $(n,d)=(5,0)$ and $(8,1)$, two stable branches remain throughout the physical parameter range examined.

The one-sided limits have different long-time behavior because one consensus state remains absorbing. Close to this boundary, the majority-rule drift is linear in the minority fraction, while the reversal correction enters at order $x^{n-d}$. The resulting consensus time is $\tau_{n,d}^{(+)}
=
n^{-1}\ln N
+
O(1).$ Changing $d$ or $\epsilon_{\uparrow}$ affects the transient part of the trajectory and the additive contribution, but not the coefficient of $\ln N$. MC first-passage measurements up to $N=10^6$ recover this logarithmic slope. The remaining offset from the deterministic estimate arises from finite-population fluctuations and the final extinction event.

The parameter $d$ represents the selectivity of a response to local agreement. A response confined to unanimous groups may rarely be activated in a large, nearly balanced population, even when its conditional probability is high. Admitting near-unanimous groups increases its macroscopic influence and may remove dependence on the initial state. This interpretation is limited to the statistical mechanism studied here. The model is well mixed, and $d$ is a prescribed coarse-grained parameter rather than an empirically calibrated threshold.

\section*{CRediT authorship contribution statement}
\textbf{R.~Muslim:} Main contributor, Conceptualization,  Methodology, Writing, Software, Formal analysis, Validation, Visualisation, Review \& editing. \textbf{R.~Anugraha.~NQZ:} Conceptualization, Formal analysis, Writing, Validation, funding acquisition, Supervision, Review \& editing. \textbf{Q.~G.~Haq:} Writing, Review \& editing. \textbf{F.~Nugroho:} Validation, Review \& editing. \textbf{I.~S.~Alam:} Review \& editing \textbf{E.~L.~Istiqomah:} Validation, Review \& editing. All authors have read, reviewed, and approved the publication of this paper.

\section*{Declaration of Interests}
The authors declare that they have no known competing financial interests or personal relationships that could have appeared to influence the work reported in this paper.


\section*{Acknowledgments}
\textbf{R.~Muslim} acknowledges support from the YST Program of the Asia Pacific Center for Theoretical Physics (APCTP), funded by the Science and Technology Promotion Fund and Lottery Fund of the Korean Government, and by the Management Talent Program of the National Research and Innovation Agency of Indonesia (BRIN). \textbf{R.~Anugraha~NQZ} acknowledges financial support from the Directorate of Research and Community Service, Ministry of Higher Education, Science, and Technology, through the Regular Fundamental Research Scheme under Contract Number 1262/UN1/DITLIT/Dit-Lit/PT.01.03/2026.


\appendix
\section{\label{app_lambda} Linear stability and critical-threshold}

This appendix derives the local expansion of the macroscopic drift in
Eq.~\eqref{eq:local_drift_compact_revised}, the majority-rule coefficient
$\Lambda_n$ in Eq.~\eqref{eq:lambda_n}, and the compact critical probability
in Eq.~\eqref{eq:critical_ebar_revised}. We start from the drift decomposition
in Eq.~\eqref{eq:drift_ebar_deltae} and write $c=1/2+\delta$, with
$|\delta|\ll1$. For convenience, let
$d_{\max}=\lfloor(n-1)/2\rfloor$ and
$P_n(\ell\mid c)=\binom{n}{\ell}c^\ell(1-c)^{n-\ell}$. At the mixed state, the binomial probability and its derivative are
\begin{equation}
    P_n\left(\ell\mid\frac12\right)
    =
    2^{-n}\binom{n}{\ell},
    \qquad
    \left.
    \frac{\partial P_n(\ell\mid c)}{\partial c}
    \right|_{c=1/2}
    =
    2^{1-n}(2\ell-n)\binom{n}{\ell}.    
\end{equation}
Introducing $B_{n,d}
    =
    \sum_{\ell=0}^{d}\binom{n}{\ell}$ and $S_{n,d}
    =
    \sum_{\ell=0}^{d}
    (n-2\ell)\binom{n}{\ell},$
and using
$A_{n,d}^{+}(c)=A_{n,d}^{-}(1-c)$, the activation functions satisfy
\begin{align}
     A_{n,d}^{-}\left(\frac12+\delta\right)
    -
    A_{n,d}^{+}\left(\frac12+\delta\right)
    &=
    -2^{2-n}S_{n,d}\delta+O(\delta^3),\\
    A_{n,d}^{-}\left(\frac12+\delta\right)
    +
    A_{n,d}^{+}\left(\frac12+\delta\right)
    &=
    2^{1-n}B_{n,d}+O(\delta^2).   
\end{align}
The difference is antisymmetric about $c=1/2$ and therefore contains only odd
powers of $\delta$, whereas the sum is symmetric and contains only even
powers.

We next evaluate the majority-rule contribution
\begin{align}
    M_n(c)
    ={}&
    \sum_{\ell=\lfloor n/2\rfloor+1}^{n-1}
    (n-\ell)\binom{n}{\ell}
    c^\ell(1-c)^{n-\ell}-
    \sum_{\ell=1}^{\lfloor(n-1)/2\rfloor}
    \ell\binom{n}{\ell}
    c^\ell(1-c)^{n-\ell}.
    \label{eq:MR_term_revised}
\end{align}
Because
$M_n(1-c)=-M_n(c)$, its local expansion can be written as
$M_n(1/2+\delta)=\Lambda_n\delta+O(\delta^3)$, where
$\Lambda_n=\left.dM_n/dc\right|_{c=1/2}$. The derivative of the elementary
factor at the mixed state is
$\left.d[c^\ell(1-c)^{n-\ell}]/dc\right|_{c=1/2}
=(2\ell-n)2^{1-n}$. Substitution into
Eq.~\eqref{eq:MR_term_revised} gives two symmetry-related contributions. In
the positive-majority sum, changing variables from $\ell$ to $k=n-\ell$
changes the summation range to $k=1,\ldots,d_{\max}$ and gives
$2^{1-n}\sum_{k=1}^{d_{\max}}k(n-2k)\binom{n}{k}$. Relabeling
$\ell\to k$ in the negative-majority sum produces the same contribution.
Therefore,
\begin{align}
    \Lambda_n
    &=
    2^{1-n}
    \left[
        \sum_{\ell=\lfloor n/2\rfloor+1}^{n-1}
        (n-\ell)(2\ell-n)\binom{n}{\ell}
        -
        \sum_{\ell=1}^{d_{\max}}
        \ell(2\ell-n)\binom{n}{\ell}
    \right]=
    2^{2-n}
    \sum_{k=1}^{d_{\max}}
    k(n-2k)\binom{n}{k}.
    \label{eq:appendix_lambda_final}
\end{align}
This reproduces Eq.~\eqref{eq:lambda_n}. Since
$1\leq k\leq d_{\max}$ implies $n-2k>0$, every term in the final sum is
positive, and hence $\Lambda_n>0$.

Combining the expansions of the majority-rule and collective-reversal
contributions in Eq.~\eqref{eq:drift_ebar_deltae} yields
\begin{align}
    v_{n,d}\left(\frac12+\delta\right)
    &=
    n2^{1-n}B_{n,d}\Delta\epsilon
    +
    \left[
        \Lambda_n
        -
        n2^{2-n}S_{n,d}\bar{\epsilon}
    \right]\delta
    +
    O(\delta^2)=
    n2^{1-n}B_{n,d}\Delta\epsilon
    +
    \lambda_{n,d}(\bar{\epsilon})\delta
    +
    O(\delta^2),
    \label{eq:appendix_local_drift}
\end{align}
where
$\lambda_{n,d}(\bar{\epsilon})
=\Lambda_n-n2^{2-n}S_{n,d}\bar{\epsilon}$. This is the local drift quoted in
Eq.~\eqref{eq:local_drift_compact_revised}. For symmetric reversal,
$\Delta\epsilon=0$, the complete drift is antisymmetric about the mixed state,
and the remainder in Eq.~\eqref{eq:appendix_local_drift} begins at
$O(\delta^3)$. The symmetric mixed state changes linear stability when
$\lambda_{n,d}(\bar{\epsilon}_c)=0$. Substitution of
Eq.~\eqref{eq:appendix_lambda_final} first gives
\begin{equation}
        \bar{\epsilon}_c(n,d)
    =
    \frac{
        \sum_{k=1}^{d_{\max}}
        k(n-2k)\binom{n}{k}
    }{
        n\sum_{\ell=0}^{d}
        (n-2\ell)\binom{n}{\ell}
    }.
    \label{eq:critical_general_appendix}
\end{equation}
The denominator can be simplified using the telescoping identity
$(n-2\ell)\binom{n}{\ell}
=n[\binom{n-1}{\ell}-\binom{n-1}{\ell-1}]$, where
$\binom{n-1}{-1}=0$. Summing from $\ell=0$ to $d$ then gives
$S_{n,d}=n\binom{n-1}{d}$.

For the numerator, we use
$k\binom{n}{k}=n\binom{n-1}{k-1}$ and
$k(k-1)\binom{n}{k}=n(n-1)\binom{n-2}{k-2}$. Evaluating the resulting partial
binomial sums up to $d_{\max}$ gives $    \sum_{k=1}^{d_{\max}}
    k(n-2k)\binom{n}{k}
    =
    \frac{n}{2}
    \left[
        n\binom{n-1}{d_{\max}}
        -
        2^{n-1}
    \right].$
Substitution of these two finite-sum identities into the critical condition
finally yields
\begin{align}
    \bar{\epsilon}_c(n,d)
    &=
    \frac{
        \sum_{k=1}^{d_{\max}}
        k(n-2k)\binom{n}{k}
    }{
        n\sum_{\ell=0}^{d}
        (n-2\ell)\binom{n}{\ell}
    }=
    \frac{
        n\binom{n-1}{d_{\max}}-2^{n-1}
    }{
        2n\binom{n-1}{d}
    },
    \label{eq:appendix_critical_compact}
\end{align}
which is the compact critical probability reported in
Eq.~\eqref{eq:critical_ebar_revised}.

We next derive the large-group behavior of the critical probability when the
dissent tolerance $d$ remains fixed as $n\to\infty$. Starting from
Eq.~\eqref{eq:appendix_critical_compact}, the critical probability can be
separated into
\begin{equation}
       \bar{\epsilon}_c(n,d)
    =
    \frac{
        \binom{n-1}{d_{\max}}
    }{
        2\binom{n-1}{d}
    }
    -
    \frac{
        2^{n-2}
    }{
        n\binom{n-1}{d}
    }. 
\end{equation}
Since $d_{\max}=\lfloor(n-1)/2\rfloor$, the binomial coefficient in the first
numerator is central, or one of the two equivalent central coefficients,
depending on the parity of $n$. Its large-$n$ behavior is $    \binom{n-1}{d_{\max}}
    =
    2^{n-1}
    \sqrt{2/\pi n}
    \left[1+O\left(n^{-1}\right)\right].$ For fixed $d$, the binomial coefficient in the denominator grows only
algebraically and satisfies $    \binom{n-1}{d}
    =
    (n^d/d!)
    \left[1+O\left(n^{-1}\right)\right].$ Substituting these approximations into the two contributions gives
\begin{align}
       \bar{\epsilon}_c(n,d)
    & \sim 2^{n-2}d!
    \sqrt{\frac{2}{\pi}}\,
    n^{-(d+1/2)} - 2^{n-2}d!\,
    n^{-(d+1)} =
    2^{n-2}d!
    \sqrt{\frac{2}{\pi}}\,
    n^{-(d+1/2)}
    \left[
        1-\sqrt{\frac{\pi}{2n}}
    \right].
\end{align}
Retaining only the leading contribution finally yields
\begin{equation}
     \bar{\epsilon}_c(n,d)
    \sim
    2^{n-2}d!
    \sqrt{\frac{2}{\pi}}\,
    n^{-(d+1/2)},
    \label{eq:critical_leading}
\end{equation}
which is the asymptotic result reported in
Eq.~\eqref{eq:critical_fixed_d_asymptotic}.


We now determine how the minimum accessible dissent tolerance grows with the
interaction size. By definition, $d_{\mathrm{acc}}(n)$ is the smallest integer
$d$ for which $\bar{\epsilon}_c(n,d)\leq1$. Using the critical
probability in Eq.~\eqref{eq:appendix_critical_compact}, the accessibility
condition can be written as
\begin{equation}
        \binom{n-1}{d}
    \geq
    \frac{
        n\binom{n-1}{d_{\max}}-2^{n-1}
    }{
        2n
    }.
\end{equation}
To analyze this condition for large $n$, let $q=n-1$ and
$r=\lfloor q/2\rfloor=d_{\max}$. The right-hand side can then be expressed as $    \frac12\binom{q}{r}
    \left[
        1-
        2^q/
            (n\binom{q}{r})
    \right].$
Using the central-binomial approximation
$\binom{q}{r}\sim2^q\sqrt{2/(\pi q)}$, the correction inside the square
brackets satisfies $2^q/(
        n\binom{q}{r})
    \sim
    \sqrt{\pi q/2n^2}
    =
    O\left(n^{-1/2}\right).$
Therefore, to leading order, the accessibility boundary is determined by $ 
        \binom{q}{d_{\mathrm{acc}}}/
        \binom{q}{r}
    =
    \frac12
    \left[
        1+O\left(q^{-1/2}\right)
    \right].$

Because the required tolerance lies below the center of the binomial
distribution, we write $d_{\mathrm{acc}}=r-s$, where $s>0$. For
$s=O(\sqrt{q})$, the local Gaussian approximation to the binomial coefficient
gives
\begin{equation}
        \frac{
        \binom{q}{r-s}
    }{
        \binom{q}{r}
    }
    =
    \exp\left(
        -\frac{2s^2}{q}
    \right)
    \left[
        1+O\left(q^{-1/2}\right)
    \right].
\end{equation}
Equating this expression to the accessibility condition and taking the
logarithm yields $ -2s^2/q
    =
    -\ln2
    +
    O\left(q^{-1/2}\right).$
It follows that
\begin{equation}
    s^2
    =
    \frac{q\ln2}{2}
    +
    O\left(\sqrt{q}\right),
    \qquad
    s
    =
    \sqrt{\frac{q\ln2}{2}}
    +
    O(1).
\end{equation}
The $O(1)$ correction also accounts for the integer-valued character of
$d_{\mathrm{acc}}$. Recalling that $q=n-1$ and
$d_{\mathrm{acc}}=d_{\max}-s$, we finally obtain
\begin{equation}
       d_{\mathrm{acc}}(n)
    =
    d_{\max}
    -
    \sqrt{\frac{(n-1)\ln2}{2}}
    +
    O(1) \simeq     \frac{n}{2}
    -
    \sqrt{\frac{n\ln2}{2}}, 
    \label{eq:critical_dcc}
\end{equation}
which is Eq.~\eqref{eq:d_accessibility_asymptotic} in the main text. 
Equivalently, the minimum aligned fraction required to activate reversal
behaves as
$1-d_{\mathrm{acc}}/n
\simeq 1/2+\sqrt{\ln2/(2n)}$.

\section{\label{app_saddle_node}Derivation of the saddle-node boundary}

This appendix derives the parametric saddle-node boundary given in
Eqs.~\eqref{eq:saddle_node_parametric_1} - \eqref{eq:saddle_node_parametric}. For fixed $n$ and $d$, we write the
macroscopic drift as $    v_{n,d}(c)
    =
    M_n(c)
    +
    n\bar{\epsilon}D_{n,d}(c)
    +
    n\Delta\epsilon Q_{n,d}(c),$ where
$D_{n,d}(c)=A_{n,d}^{-}(c)-A_{n,d}^{+}(c)$ and
$Q_{n,d}(c)=A_{n,d}^{-}(c)+A_{n,d}^{+}(c)$. A generic saddle-node
bifurcation occurs when a stable and an unstable fixed point merge into a
double root at $c=c_{\mathrm{sn}}$. The corresponding conditions are
$v_{n,d}(c_{\mathrm{sn}})=0$ and
$v_{n,d}'(c_{\mathrm{sn}})=0$.

Because the drift is linear in $\bar{\epsilon}$ and $\Delta\epsilon$, these
conditions form a linear system for the two reversal parameters. Suppressing
the subscripts and evaluating all functions at $c=c_{\mathrm{sn}}$, the system
can be written as
\begin{equation}
     \begin{pmatrix}
        D & Q\\
        D' & Q'
    \end{pmatrix}
    \begin{pmatrix}
        \bar{\epsilon}_{\mathrm{sn}}\\
        \Delta\epsilon_{\mathrm{sn}}
    \end{pmatrix}
    =
    -\frac{1}{n}
    \begin{pmatrix}
        M\\
        M'
    \end{pmatrix}.
\end{equation}
The determinant of the coefficient matrix is
$\mathcal{J}_{n,d}(c)=D_{n,d}(c)Q_{n,d}'(c)
-D_{n,d}'(c)Q_{n,d}(c)$. Provided that
$\mathcal{J}_{n,d}(c_{\mathrm{sn}})\neq0$, the two reversal parameters are
obtained directly using Cramer's rule:
\begin{align}
    \bar{\epsilon}_{\mathrm{sn}}(c)
    &=
    \frac{
        (-M_n/n)Q_{n,d}'
        -
        Q_{n,d}(-M_n'/n)
    }{
        D_{n,d}Q_{n,d}'-D_{n,d}'Q_{n,d}
    }
    =
    \frac{
        Q_{n,d}M_n'-M_nQ_{n,d}'
    }{
        n\mathcal{J}_{n,d}
    },\\
    \Delta\epsilon_{\mathrm{sn}}(c)
    &=
    \frac{
        D_{n,d}(-M_n'/n)
        -
        D_{n,d}'(-M_n/n)
    }{
        D_{n,d}Q_{n,d}'-D_{n,d}'Q_{n,d}
    }
    =
    \frac{
        M_nD_{n,d}'-D_{n,d}M_n'
    }{
        n\mathcal{J}_{n,d}
    }.
\end{align}
Restoring the explicit dependence on $c$, we recover
\begin{align}
    \bar{\epsilon}_{\mathrm{sn}}(c)
    &=
    \frac{
        Q_{n,d}(c)M_n'(c)
        -
        M_n(c)Q_{n,d}'(c)
    }{
        n\mathcal{J}_{n,d}(c)
    },\\
    \Delta\epsilon_{\mathrm{sn}}(c)
    &=
    \frac{
        M_n(c)D_{n,d}'(c)
        -
        D_{n,d}(c)M_n'(c)
    }{
        n\mathcal{J}_{n,d}(c)
    },
\end{align}
which are Eqs.~\eqref{eq:saddle_node_parametric_1} - \eqref{eq:saddle_node_parametric}. As $c$ is varied over
$0<c<1$, these expressions parametrically generate the double-root boundary.
Only points satisfying
$0\leq\bar{\epsilon}_{\mathrm{sn}}\leq1$ and
$|\Delta\epsilon_{\mathrm{sn}}|
\leq\min(\bar{\epsilon}_{\mathrm{sn}},
1-\bar{\epsilon}_{\mathrm{sn}})$ belong to the physical probability domain.

The reflection symmetry of the phase diagram follows directly from
$M_n(1-c)=-M_n(c)$,
$D_{n,d}(1-c)=-D_{n,d}(c)$, and
$Q_{n,d}(1-c)=Q_{n,d}(c)$. Substitution into the parametric solution gives
$\bar{\epsilon}_{\mathrm{sn}}(1-c)
=\bar{\epsilon}_{\mathrm{sn}}(c)$ and
$\Delta\epsilon_{\mathrm{sn}}(1-c)
=-\Delta\epsilon_{\mathrm{sn}}(c)$. The two branches are therefore mirror
images about $\Delta\epsilon=0$. At the symmetric point $c=1/2$, one has
$M_n(1/2)=D_{n,d}(1/2)=Q_{n,d}'(1/2)=0$,
$M_n'(1/2)=\Lambda_n$, and
$D_{n,d}'(1/2)=-2^{2-n}S_{n,d}$. The parametric solution consequently reduces
to
\begin{align}
 \Delta\epsilon_{\mathrm{sn}}\left(\frac12\right)=0,
    \quad
    \bar{\epsilon}_{\mathrm{sn}}\left(\frac12\right)
    =
    -\frac{\Lambda_n}{
        nD_{n,d}'(1/2)
    }
    =
    \frac{\Lambda_n}{
        n2^{2-n}S_{n,d}
    }
    =
    \bar{\epsilon}_c(n,d).   
\end{align}
Thus, the two symmetry-related saddle-node boundaries meet the symmetric line
at the mixed-state stability threshold.

\section{\label{app_stationary_branches}Stationary branches for $n=5$}

This appendix derives the stationary ordered branches given in
Eqs.~\eqref{eq:m5_d0} and \eqref{eq:m5_d1}. We consider symmetric reversal,
$\epsilon_{\uparrow}=\epsilon_{\downarrow}=\epsilon$, and introduce the
magnetization $m=2c-1$, so that $c=(1+m)/2$. Because
$dm/d\tau=2\,dc/d\tau$, the factor of two does not affect the locations or
linear stability of the stationary solutions. We therefore express the drift
$v_{5,d}(c)$ directly as a function of $m$. For $n=5$, the pure majority-rule drift in
Eq.~\eqref{eq:MR_term_revised} contains positive-majority configurations with
$\ell=3$ and $4$, and negative-majority configurations with $\ell=1$ and $2$.
It therefore reduces to

\begin{align}
    M_5(c)
    =&
    (5-3)\binom{5}{3}c^3(1-c)^2
    +
    (5-4)\binom{5}{4}c^4(1-c)-
    1\binom{5}{1}c(1-c)^4
    -
    2\binom{5}{2}c^2(1-c)^3\nonumber \\
    =&    20c^3(1-c)^2
    +
    5c^4(1-c)
    -
    5c(1-c)^4
    -
    20c^2(1-c)^3.
\end{align}
Substitution of $c=(1+m)/2$ and collection of equal powers of $m$ gives $ M_5(m)
    =
    5m/16
    \left(
        7-10m^2+3m^4
    \right).$

For strict-unanimity activation, $d=0$, the negative and positive activation
sectors contain only the configurations $\ell=0$ and $\ell=5$, respectively.
Using the binomial group-composition probability, one obtains $A_{5,0}^{-}(c)=
    P_5(0\mid c)
    =
    (1-c)^5$ and $
    A_{5,0}^{+}(c)=
    P_5(5\mid c)
    =
    c^5.$
Therefore, $A_{5,0}^{-}(c)-A_{5,0}^{+}(c)= (1-c)^5-c^5.$
Substituting $c=(1+m)/2$ gives

\begin{align}
    A_{5,0}^{-}(m)-A_{5,0}^{+}(m)
=
    \frac{(1-m)^5-(1+m)^5}{32} = -\frac{m}{16}
    \left(
        m^4+10m^2+5
    \right).
\end{align}
The symmetric drift is consequently
\begin{align}
    v_{5,0}(m)
    &=
    M_5(m)
    +
    5\epsilon
    \left[
        A_{5,0}^{-}(m)-A_{5,0}^{+}(m)
    \right]=
    -\frac{5}{16}m
    \left[
        (\epsilon-3)m^4
        +
        10(1+\epsilon)m^2
        +
        5\epsilon-7
    \right].
\end{align}
The stationary condition $v_{5,0}(m^\ast)=0$ always admits the mixed solution
$m^\ast=0$. Defining $y=m^{\ast2}$ reduces this quartic equation to $(\epsilon-3)y^2
    +
    10(1+\epsilon)y
    +
    5\epsilon-7
    =
    0.$
The solution with the positive sign in front of the square root lies outside
the physical interval $0\leq y\leq1$. The physical nonzero solution is
therefore
\begin{equation}
    y^\ast
    =
    \frac{
        5(1+\epsilon)
        -
        2\sqrt{1+18\epsilon+5\epsilon^2}
    }{
        3-\epsilon
    }, \quad \text{or} \quad     m_\pm^\ast(\epsilon)
    =
    \pm
    \left[
        \frac{
            5(1+\epsilon)
            -
            2\sqrt{1+18\epsilon+5\epsilon^2}
        }{
            3-\epsilon
        }
    \right]^{1/2},
    \label{eq:mag_app_1}
\end{equation}
which is Eq.~\eqref{eq:m5_d0}.

Similarly, for $d=1$, the reversal-eligible sectors contain both unanimous
groups and groups with one dissenter. After substituting
$c=(1+m)/2$, the difference between the corresponding activation
probabilities becomes
$A_{5,1}^{-}(m)-A_{5,1}^{+}(m)=m(m^4-5)/4$. Combining this result with the
pure majority-rule contribution yields
\begin{align}
    v_{5,1}(m)
    =
    M_5(m)
    +
    5\epsilon
    \left[
        A_{5,1}^{-}(m)-A_{5,1}^{+}(m)
    \right]=
    \frac{5}{16}m
    \left[
        (3+4\epsilon)m^4
        -10m^2
        +7-20\epsilon
    \right].
\end{align}
The stationary condition $v_{5,1}(m^\ast)=0$ always admits the mixed solution
$m^\ast=0$. For the nonzero branches, setting $y=m^{\ast 2}$ reduces the
remaining equation to a quadratic equation in $y$. The physical root is
$y^\ast=[5-2\sqrt{1+8\epsilon+20\epsilon^2}]/(3+4\epsilon)$, and therefore
\begin{equation}
    m_\pm^\ast(\epsilon)
    =
    \pm
    \left[
        \frac{
            5-2\sqrt{1+8\epsilon+20\epsilon^2}
        }{
            3+4\epsilon
        }
    \right]^{1/2}.
    \label{eq:mag_app_2}
\end{equation}
This reproduces Eq.~\eqref{eq:m5_d1} in the main text.

\section{\label{app_consensus_time}
Boundary-dominated estimate of the consensus time toward
\texorpdfstring{$c=1$}{c=1}}

This appendix derives the boundary approximation in
Eq.~\eqref{eq:tau_plus_boundary} and its large-$N$ scaling in
Eq.~\eqref{eq:tau_plus_scaling}. We consider the one-sided case
$\epsilon_{\downarrow}=0$, for which the all-$+1$ configuration is absorbing,
and introduce the minority concentration $x=1-c$. Consensus corresponds to
$x=0$, while the finite-population cutoff $c_f=1-1/N$ corresponds to
$x_f=1/N$. Near the absorbing boundary, the negative-majority activation probability can
be written as
\begin{equation}
       A_{n,d}^{-}(1-x)
    =
    \sum_{\ell=0}^{d}
    \binom{n}{\ell}
    (1-x)^\ell x^{n-\ell}. 
\end{equation}
As $x\rightarrow0$, the leading contribution comes from the largest eligible
value $\ell=d$, because this term contains the smallest power of $x$.
Consequently,
$A_{n,d}^{-}(1-x)=\binom{n}{d}x^{n-d}
+O(x^{n-d+1})$. The majority-rule contribution has the corresponding
boundary expansion $M_n(1-x)=nx+O(x^2)$. Substitution into the one-sided drift
therefore gives $v_{n,d}^{(+)}(1-x)
    =
    nx
    +
    n\epsilon_{\uparrow}
    \binom{n}{d}x^{n-d}
    +
    O(x^2)
    +
    O(x^{n-d+1}).$
Defining $p=n-d-1$ and
$a_{n,d}^{(+)}=\epsilon_{\uparrow}\binom{n}{d}$, the retained boundary terms
take the compact form as
\begin{equation}
    v_{n,d}^{(+)}(1-x)
    \simeq
    nx\left[
        1+a_{n,d}^{(+)}x^p
    \right].
    \label{eq:appendix_boundary_drift}
\end{equation}
The admissible range
$0\leq d\leq\lfloor(n-1)/2\rfloor$ ensures that $p\geq1$.

The deterministic travel time from $c_0$ to the finite-size boundary
$1-1/N$ is obtained by integrating the inverse drift. After changing variables
from $c$ to $x=1-c$, Eq.~\eqref{eq:appendix_boundary_drift} gives
\begin{align}
    \tau_{n,d}^{(+)}
    \simeq&
    \frac{1}{n}
    \int_{1/N}^{1-c_0}
    \frac{dx}{
        x\left[1+a_{n,d}^{(+)}x^p\right]
    } =
    \frac{1}{n}
    \int_{1/N}^{1-c_0} \left[ \frac{1}{x}
    -
    \frac{a x^{p-1}}{1+a x^p} \right] dx \nonumber \\
    \simeq&
    \frac{1}{n}
    \left[
        \ln\!\bigl(N(1-c_0)\bigr)
        -
        \frac{1}{p}
        \ln\!\left(
            \frac{
                1+a_{n,d}^{(+)}(1-c_0)^p
            }{
                1+a_{n,d}^{(+)}N^{-p}
            }
        \right)
    \right],
    \label{eq:appendix_tau_plus_boundary}
\end{align}
which reproduces Eq.~\eqref{eq:tau_plus_boundary}. For $d=0$, one has
$p=n-1$ and $a_{n,0}^{(+)}=\epsilon_{\uparrow}$, so the result reduces to the
strict-unanimity expression.

The large-population behavior follows from
$\ln[1+a_{n,d}^{(+)}N^{-p}]
=a_{n,d}^{(+)}N^{-p}+O(N^{-2p})$. Hence,
Eq.~\eqref{eq:appendix_tau_plus_boundary} becomes
\begin{equation}
    \tau_{n,d}^{(+)}
    =
    \frac{1}{n}\ln N
    +
    C_{n,d}^{(+)}
    (c_0,\epsilon_{\uparrow})
    +
    O(N^{-p}),
    \label{eq:appendix_tau_plus_scaling}
\end{equation}
where, within the boundary approximation,
\begin{equation}
      C_{n,d}^{(+)}
    (c_0,\epsilon_{\uparrow})
    =
    \frac{1}{n}
    \left[
        \ln(1-c_0)
        -
        \frac{1}{p}
        \ln\!\left(
            1+a_{n,d}^{(+)}(1-c_0)^p
        \right)
    \right].  
\end{equation}
Since $p\geq1$, the correction $O(N^{-p})$ vanishes as $N\rightarrow\infty$,
and Eq.~\eqref{eq:appendix_tau_plus_scaling} recovers the logarithmic scaling
reported in Eq.~\eqref{eq:tau_plus_scaling}. The coefficient $1/n$ originates
from the linear majority-rule drift $v_{n,d}^{(+)}(1-x)\sim nx$ and is
therefore independent of both $d$ and $\epsilon_{\uparrow}$. These parameters
affect only the finite, $N$-independent contribution within the present
approximation.

Equation~\eqref{eq:appendix_tau_plus_boundary} retains the leading
reversal-dependent boundary correction but neglects higher-order terms in the
majority-rule drift. When the complete drift is integrated from an initial
condition that is not asymptotically close to $c=1$, those terms can modify
the additive constant $C_{n,d}^{(+)}$ without changing the leading
$(1/n)\ln N$ dependence. The corresponding result for consensus toward
$c=0$ follows from the exchange
$c\leftrightarrow1-c$ together with
$\epsilon_{\uparrow}\leftrightarrow\epsilon_{\downarrow}$.

\bibliographystyle{elsarticle-num}

\bibliography{cas-refs}

\vskip3pt
\end{document}